\RequirePackage{fix-cm}
\documentclass[smallextended]{svjour3}       
\smartqed  
\usepackage{graphicx}
\usepackage{amsmath}
\usepackage{amsfonts}
\usepackage{natbib}
\usepackage{array}
\usepackage{xcolor}
\usepackage{float}
\newcolumntype{L}[1]{>{\raggedright\let\newline\\\arraybackslash\hspace{0pt}}m{#1}}
\newcolumntype{C}[1]{>{\centering\let\newline\\\arraybackslash\hspace{0pt}}m{#1}}
\newcolumntype{R}[1]{>{\raggedleft\let\newline\\\arraybackslash\hspace{0pt}}m{#1}}
\journalname{Transport in Porous Media}
\begin{document}

\title{Lagrangian complexity persists with multimodal flow forcing in poroelastic systems}
%

\subtitle{}

\titlerunning{Multimodal spectral forcing enhances Lagrangian complexity}        

\author{M. G. Trefry        \and
        D. R. Lester        \and
        G. Metcalfe         \and
        J. Wu
}


\institute{M. G. Trefry \at
           Floreat, Western Australia, Australia \\
           \email{michaelgtrefry@gmail.com}
           \and
           D. R. Lester \at
           School of Engineering, RMIT, Melbourne, Victoria, Australia \\   \and
           G. Metcalfe \at
           School of Engineering, Swinburne University of Technology, Hawthorn, Victoria, Australia \\
           \and
           J. Wu \at
           School of Engineering, RMIT, Melbourne, Victoria, Australia \\
}

\date{Received: date / Accepted: date}

\maketitle

\begin{abstract}
We extend previous analyses of the origins of complex transport dynamics in poroelastic media to  the case where the input transient signal at a boundary is generated by a multimodal spectrum. By adding harmonic and anharmonic modal frequencies as perturbations to a fundamental mode we examine how such multimodal signals affect the Lagrangian complexity of poroelastic flow.  While the results apply to all poroelastic media (industrial, biological and geophysical), for concreteness we couch the discussion in terms of unpumped coastal groundwater systems having a discharge boundary forced by tides. Particular local regions of the conductivity field generate saddles that hold up and braid (mix) trajectories, resulting in unexpected behaviours of groundwater residence time distributions and topological mixing manifolds near the tidal boundary. While increasing spectral complexity can reduce the occurrence of periodic points, especially for anharmonic spectra with long characteristic periods, other signatures of Lagrangian complexity persist. The action of natural multimodal tidal signals on confined  groundwater flow in heterogeneous aquifers can induce exotic flow topologies and mixing effects that are profoundly different to conventional concepts of groundwater discharge processes. Taken together, our results imply that increasing spectral complexity results in more complex Lagrangian structure in flows through poroelastic media. 
\end{abstract}

\section{Introduction}
\label{intro}

It has long been recognized that unsteady groundwater flows admit a range of behaviours that are critical to the transport and fate of dissolved contaminants in the subsurface \citep{weeks:1998aa}. Understanding these behaviours at the representative elemental volume scale is a topic of continuing interest. Due to the unsteady nature of these flows, it is difficult to detect and understand the corresponding transport structures in the Eulerian frame. The Lagrangian approach is of more relevance here as it provides a convenient framework for detecting and classifying kinematic features that can control fluid mixing, segregation, discharge (and even chaotic) phenomena in porous media in the low Reynolds number regime. Together these kinds of transport phenomena describe \textit{complex Lagrangian structures} which engender potentially profound impacts on solute migration and reaction \citep{Toroczkai:1998aa,Tel2005,Valocchi:2019aa}. Building on early conceptual work by \citet{Ottino:1989aa}, \citet{Sposito:2001aa,Sposito:2006aa} and others \citep{Lester:2009ab,Lester:2010aa,Metcalfe:2010aa,Trefry:2012aa,Mays:2012,Lester:2015aa}, groundwater researchers are now able to predict, observe and engineer complex Lagrangian structures in saturated porous media at the laboratory \citep{Zhang:2009aa,Metcalfe:2010ab} and field \citep{Cho:2019aa} scales, with quantified benefits for mixing and reactivity enhancement.

Complex Lagrangian structures have also been predicted to occur in natural groundwater environments, i.e. Darcian systems where engineered pumping activity is absent. Using a conventional Darcian confined flow model in two dimensions (vertically averaged), \citet{Trefry:2019aa} showed computationally that complex Lagrangian structures arose where aquifer heterogeneity and compressibility were combined with sinusoidal boundary forcing. This analysis was extended by \citet{Wu:2019aa} to consider wide ranges of parameter values, highlighting that the spectra of natural ocean tides around the globe provide significant potential to induce Lagrangian complexity in coastal aquifer flows; other important hydraulic parameters are the local aquifer matrix properties of compressibility, diffusivity and heterogeneity. However the time-periodic spectral boundary condition imposed on the discharging groundwater flow enabled \citet{Trefry:2019aa} to utilize concepts from time-periodic maps, such as periodic points and manifolds to identify and describe the Lagrangian structures.  These structures, quantified by the local net fluid deformation over one period of the trajectory, organize a rich set of transport structures that set fluid residence times, segregation, and mixing. We note that since the basic ingredients of porous media heterogeneity, compressibility and time-dependent forcing also arise in a number of other geophysical, industrial and biological systems, these systems may also exhibit complex transport dynamics.

Motivated by coastal/lacustrine/riverine aquifers discharging to receiving surface water bodies, \citet{Trefry:2019aa} and \citet{Wu:2019aa} employed a simple transient formulation of the forcing boundary condition in terms of a single sinusoidal mode $\omega = 2 \pi f$ where $f$ is the frequency of the mode. This single mode approximation is a common and useful approach for theoretical analysis of tidally affected groundwater systems governed by linear dynamics, but it does not capture the temporal complexity of natural discharge boundary signals in hydrological systems. Natural boundaries often display frequency spectra that contain several or many frequency modes with different amplitudes and phases \citep{Munk:1966aa}, for example tidal stage heights, or seasonal rainfall profiles. Given that the previous single-mode studies have predicted such remarkable kinematic and transport behaviours, it is logical to ask what effect a multimode forcing spectrum may have on the Lagrangian topology of  flow in poroelastic media. This question is the subject of this paper, and which we will pursue via a model tidally influenced groundwater discharge problem.

In the remainder of this paper we focus on answering the key question on what impact increasing complexity in the forcing spectrum has on the Lagrangian structure of discharging groundwater flows and, by extension, to any open poroelastic flows. In the next section we begin by briefly considering the formal mathematical periodicities of multimodal spectra, before introducing our archetypal poroelastic groundwater discharge problem, as employed in the earlier works but modified slightly for the present multimodal context. The various Lagrangian metrics to be used in this analysis are introduced, and we then develop a set of numerical experiments to test the effect of increasing spectral complexity on Lagrangian properties of the flow solutions, before finishing with a discussion of findings and ramifications for studies of discharging flow and transport in poroelastic systems.

\section{Periodicity in multimodal spectra}
\label{sec:1}

It is useful to consider some fundamental aspects of the spectral representation of a time-dependent signal $F(t)$ where $t$ is time. We focus on the associated spectral representation $\mathcal{F}(\omega)$ defined by the Fourier transformation of $F$
\begin{eqnarray}\label{eq:fouriertransform}
  \mathcal{F}(\omega) \equiv \frac{1}{\sqrt{2 \pi}} \int_{- \infty}^{\infty} F(t) e^{i \omega t} dt.
\end{eqnarray}
Equation (\ref{eq:fouriertransform}) generates the complex-valued $\mathcal{F}$ function which contains amplitude ($|\mathcal{F}|$) and phase ($\textrm{arg}\,\mathcal{F}$) information for every angular frequency $\omega$ present in the spectrum of $F$. Where $F$ is periodic, i.e. there exists a real number $P$ (the period) that satisfies $F(t+P) = F(t)$ for all values of $t$, the resulting spectrum $\mathcal{F}$ consists of a discrete (but possibly infinite) set of modes $\omega_{m}$ for $m = 1,2,3$ and so on. If $F$ is not periodic in time then $\mathcal{F}$ is a dense function of $\omega$, i.e. $\mathcal{F} \neq 0$ over continuous intervals (possibly infinite) of $\omega$.
\subsection{Commensurable spectra and characteristic period $P$}
\label{sec:2}
If we confine attention to the context of natural periodically forced groundwater systems we can limit our analysis to periodic boundary conditions characterized by a finite set of $M$ spectral modes $\omega_{m}$ for $m = 1,2, ... M$, termed an $M$-spectrum. Anticipating the use of dynamical systems tools to classify repeating Lagrangian structures, including Poincar\'e sections, periodic points etc, we must focus on the periodicity characteristics displayed by $M$-spectra. It is not immediately apparent, but signals constructed from simple finite sums of (inherently periodic) sinusoidal functions may display infinitely many different periods, or even be inherently non-periodic. Where each pair of modes $\omega_{m},\omega_{n}$ can be related by $\omega_{m} / \omega_{n} = p / q$ for $p,q$ positive integers such that $p \neq q$, i.e. the modes are distinct and rationally related, the spectrum is termed \textit{commensurable} and an infinite number of exact periods exist for $F$. By convention we choose the smallest of these periods as the \textit{characteristic period} $P$ for $F$ (see Appendix \ref{app:period} for further discussion). Where the ratio of any two particular modes within the spectrum cannot be expressed exactly as a rational number, the spectrum is \textit{incommensurable} and no exact period exists for $F$, thereby precluding identification of exact periodic points for analysis. However, as most hydrological applications involve boundary conditions corresponding to commensurable spectra---since the constituent modes are calculated from head or flux values expressed digitally as rational values---periodic points may formally exist in the derived flows. 

An important consequence of the presence of commensurable $M$-spectra on the analysis of groundwater transport is that the characteristic period $P$ now may be significantly longer than the periods of the individual modes. Previous studies \citep{Trefry:2019aa,Wu:2019aa} utilised periodicity of the forcing signal (and thus the associated fluid velocity field) to resolve and understand Lagrangian transport structures via dynamical systems tools and techniques such as Poincar\'{e} sections, stability of periodic points, stable and unstable manifolds, etc. The applicability of these techniques hinged on the fact that typical particle trajectories would remain within the aquifer domain over many flow periods, and so the finite-time, intra-period dynamics only played a secondary role in controlling transport. Conversely, as the flow period becomes longer, the intra-period transport dynamics become  more important and the periodic Lagrangian structures only provide a skeletal, low-level description of the complete transport dynamics. In the limit of incommensurable $M$-spectra, the characteristic period $P$ effectively diverges to infinity, and only finite-time dynamics can completely describe the transport dynamics of the flow. Henceforth we confine attention to periodic boundary conditions that are defined in terms of commensurable $M$-spectra. We now turn to defining our example tidal groundwater problem.

\section{Tidal problem definition and solution method}

\begin{figure}[ht!]
    \includegraphics[width=9cm]{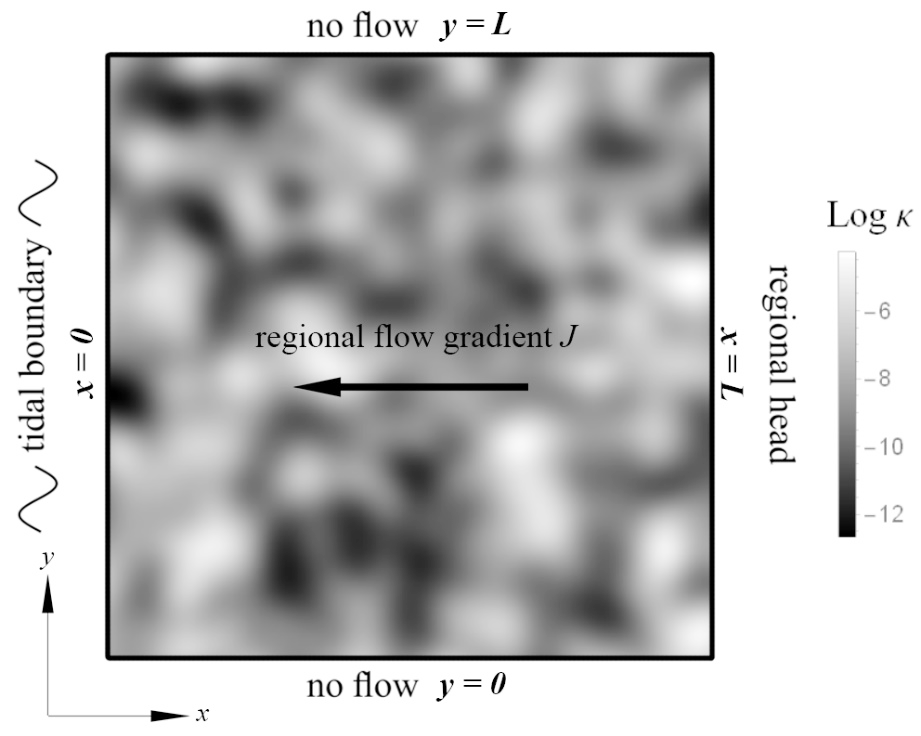}
    \caption{Schematic domain of the model coastal groundwater discharge problem (adapted from \citet{Wu:2019aa}).}
    \label{fig:domain}
\end{figure}

Here we provide an abbreviated description of our two-dimensional (in plan view) tidally forced coastal groundwater problem with regional flow. More comprehensive descriptions may be found in \citet{Trefry:2019aa} and \citet{Wu:2019aa}. In scaled coordinates the $L \times L$ coastal aquifer domain (see Figure \ref{fig:domain}) is represented by the unit square region $\textbf{x} = (x,y) \in [0,1] \times [0,1]$. We employ the (vertically integrated) linear Darcy groundwater flow model for the head $h(\mathbf{x},t)$
\begin{align}\label{eq:gwproblem}
  &S \frac{\partial h}{\partial t} = \nabla \cdot (K(\mathbf{x}) \nabla h), \\
  &\frac{\partial h}{\partial y}\Bigr|_{y=0} = \frac{\partial h}{\partial y}\Bigr|_{y=1} = 0, \; h(1,y,t) = J, \; h(0,y,t) = G(t),
\end{align}
\noindent where $S$ is the storativity of the aquifer, $K(\mathbf{x})$ is the heterogeneous hydraulic conductivity field, $J$ is the regional discharge gradient towards the tidal boundary at $x = 0$ and $G(t)$ is the tidal forcing function at the tidal boundary $x=0$. The Darcy flux vector field $\mathbf{q}(\mathbf{x},t)$ and the groundwater velocity field $\textbf{v}(\mathbf{x},t)$ are generated from the head solution $h(x,y,t)$ by the poroelastic Darcy model
\begin{align}
  \textbf{q} &= -K \nabla h,  \label{eq:fluxfield} \\
  \varphi(h) &= \varphi_{\textrm{ref}} + S \, (h - h_{\textrm{ref}}),   \label{eq:porosityfield} \\
  \textbf{v} &= \textbf{q} / \varphi,   \label{eq:velocityfield}
\end{align}
where $\varphi(\mathbf{x},t)$ is the local porosity of the aquifer, and $\varphi_{\textrm{ref}}$, $h_{\textrm{ref}}$ are the reference porosity and head used in the linear porosity equation (\ref{eq:porosityfield}). The head and velocity distributions are time-dependent due to the tidal forcing function $G(t)$, and spatially heterogeneous through the agency of the hydraulic conductivity field $K$, assumed for simplicity to be Gaussian-correlated and lognormal with mean $K_{\textrm{eff}}$, variance $\sigma^{2}_{\textrm{log} K}$ and isotropic correlation scale $\lambda$.

In this work we focus specifically on how spectral complexity of the tidal forcing signal $G(t)$ may impact the presence and/or prevalence of Lagrangian structure in nearby groundwater flows. Let us assume that the tidal forcing signal $G(t)$ can be represented by a commensurable $M$-spectrum, i.e.
\begin{eqnarray}\label{eq:Fspectrum}
  &G(t) = \textsl{g}_{0} +  \sum\limits_{m=1}^{M}{\textsl{g}_{m} \; e^{i (\omega_{m} \,
t + \theta_{m})}}.
\end{eqnarray}
By a suitable rescaling of the height datum for the problem, and without loss of generality, we can neglect the constant offset term $\textsl{g}_{0}$. Linearity of the groundwater equation (\ref{eq:gwproblem}) then ensures that the head, Darcy flux and porosity solutions take the $M$-spectral forms
\begin{eqnarray}
  &h(\textbf{x},t) = h_{0}(\textbf{x}) +  \sum\limits_{m=1}^{M}{h_{m}(\textbf{x}) \; e^{i (\omega_{m} \, t + \theta_{m})}} \label{eq:hspectrum} \\
  &\textbf{q}(\textbf{x},t) = \textbf{q}_{0}(\textbf{x}) +  \sum\limits_{m=1}^{M}{\textbf{q}_{m}(\textbf{x}) \; e^{i (\omega_{m} \, t + \theta_{m})}} \label{eq:qspectrum} \\
  &\varphi(\textbf{x},t) = \varphi_{0}(\textbf{x}) +  \sum\limits_{m=1}^{M}{\varphi_{m}(\textbf{x}) \; e^{i (\omega_{m} \, t + \theta_{m})}} \label{eq:phispectrum} 
\end{eqnarray}

\noindent The quantities $h_{0}(\mathbf{x})$, $\textbf{q}_{0}(\mathbf{x})$, $\varphi_{0}(\mathbf{x})$ indicate relevant solutions of the steady subproblem \citep{Trefry:2019aa}. Hence, assembly of the full transient velocity field $\textbf{v}(\mathbf{x},t)$ is achieved by solving numerically for the Fourier amplitudes $h_{m}(\mathbf{x})$, $\textbf{q}_{m}(\mathbf{x})$, building full head and flux solutions according to \eqref{eq:hspectrum}, \eqref{eq:qspectrum}, and then constructing first $\varphi_{m}(\mathbf{x}) = S \, h_{m}(\mathbf{x})$ then reconstituting $\varphi(\mathbf{x},t)$ according to \eqref{eq:porosityfield} and finally $\textbf{v}(\mathbf{x},t)$ according to \eqref{eq:velocityfield}. Numerical solution of the Fourier amplitudes is obtained using a multimodal generalization of the spectral method of \citet{Trefry:2009aa,Trefry:2019aa} appropriate for $M$-spectra, resulting in high-resolution flux fields that strictly satisfy continuity at all interior locations. 

Because of the multimodal nature of the problem, the dimensionless dynamical and heterogeneity parameters originally developed for the 1-spectrum problem \citep{Trefry:2019aa} now need adjustment, as is summarized for each frequency mode $m$ in Table \ref{table:nondimpars}. In multimodal cases, effective parameter values can be estimated by simple amplitude weighting of the modal parameter values. For example, the effective multimodal Townley number $\mathcal{T}_{\textrm{eff}}$ is given by
\begin{equation} \label{eq:amplitudeweight}
    \mathcal{T}_{\textrm{eff}} = \sum\limits_{m=1}^{M} \textsl{g}_{m} \mathcal{T}_{m} \Biggm/ \sum\limits_{m=1}^{M} \textsl{g}_{m}.
\end{equation}
Throughout this paper effective multimodal quantities will be understood to be evaluated using the above amplitude weighting scheme, e.g. $\mathcal{T}_{\textrm{eff}}$, $\mathcal{G}_{\textrm{eff}}$, $\mathcal{C}_{\textrm{eff}}$ etc.

\begin{table}[ht]
\caption{Dimensionless parameters for the multimodal tidal problem, with the mode-dependence made explicit. Effective parameter values may be determined using modal amplitude weighting, as described in equation \eqref{eq:amplitudeweight}.  $x_{\textrm{taz}}$ is as defined in \citet{Trefry:2019aa} using modal frequency $\omega_{m}$. }
\begin{center}\linespread{0.5}
  \begin{tabular}{|C{1.8cm} | C{2.3cm} | C{1.8cm} | C{1.0cm} | C{2.8cm}|} 
 \hline
 \small\textbf{Parameter} & \small\textbf{Name} & \small\textbf{Definition} & \small\textbf{Range} & \small\textbf{Motivation} \\ [0.5ex] 
 \hline
\multicolumn{5}{l}{\textit{Dynamical Parameter Set} $Q$} \\
\hline
 $\mathcal{T}_{m}$ & \small Townley number & $L^{2}S \omega_{m} / K_{\textrm{eff}}$ & $[0,\infty)$ & \small Ratio of diffusive time scale to modal period \\ [1.5ex]
 $\mathcal{G}_{m}$ & \small Tidal strength & $\textsl{g}_{m} / J L$ & $[0,\infty)$ & \small Ratio of modal amplitude to inland head \\ [1.5ex]
 $\mathcal{C}_{m}$ & \small Tidal compression ratio & $S \textsl{g}_{m} / \varphi_{\textrm{ref}}$ & $[0,1]$ & \small Maximum relative change in porosity due to tidal mode \\
 \hline
\multicolumn{5}{l}{\textit{Heterogeneity Parameter Set} $\chi$} \\
 \hline
  $\mathcal{H}_{t,m}$ & \small Temporal character & $\frac{\lambda \, \varphi_{\textrm{ref}} \, \omega_{m}}{2 \pi K_{\textrm{ref}} \, |J|}$ & $[0,\infty)$ & \small Ratio of Darcy drift time scale to modal period \\ [1.5ex]
 $\mathcal{H}_{x,m}$ & \small Spatial character & $x_{\textrm{taz}}/\lambda $ & $[0,L/\lambda]$ & \small Number of correlation scales in the mode's tidally active zone* \\ [1.5ex]
  $\mathcal{G}_{m} \, \sigma^{2}_{\textrm{log} K}$ & \small Vorticity or flow reversal number & $\mathcal{G}_{m} \, \sigma^{2}_{\textrm{log} K}$ & $[0,\infty)$ & \small Density of flow reversals in the mode's tidally active zone \\
  \hline
\end{tabular}
\label{table:nondimpars}
\end{center}
\end{table}

\subsection{Remarks on groundwater problem scope}
\label{sec:scope}
Having defined our model poroelastic groundwater problem, we now make some necessary comments on issues of scope for the subsequent analysis. First, we recognize that coastal aquifer processes almost invariably involve variable density flows due to the salinity differences that often exist between sea water and groundwater, yet we assume isohaline fluids in our model as defined above. Inevitably this will render inaccurate our model predictions near the discharge boundary in cases where the salinity differences are large. However, as discussed by \citet{Wu:2019aa}, penetration distances of saline wedges into the coastal aquifers are typically small when compared with the penetration of tidal forcing signals, which may persist for several kilometres or more from shorelines. It has been shown computationally that important Lagrangian structures may occur at significant distances inland from the tidal shore and well beyond the saline wedge. Therefore, it is valid to pursue the present analysis while bearing in mind that particular choices of problem parameters may entail a loss of accuracy of flow predictions near the discharge boundary. In non-coastal tidal groundwater discharge systems (riverine, lacustrine) the isohaline assumption is often more appropriate.

Secondly, we choose to consider only a two-dimensional model despite the fact that buoyancy driven flow in heterogeneous systems will be strongly three-dimensional in nature, especially near the saline wedge. \citet{Werner:2013aa} note that transient, fully three-dimensional wedge modelling is computationally difficult and still relatively rare (see, e.g. \citet{Trefry:2007aa} for a homogeneous 3D example and \citet{Mabrouk:2019aa} for a heterogeneous 3D example), so that it is currently infeasible to perform  Lagrangian analyses at the required resolution for heterogeneous buoyancy driven systems in three dimensions. In fact we contend that there is still much to understand in the two-dimensional problem before tackling three dimensions, where the indications are that Lagrangian structures will persist and be much more complex than in two dimensions \citep{Holm:1991aa,Haller2001248,Blazevski201446,Wiggins:2010aa,Lester:2015aa,Smith:2016aa,Ravu:2020aa}. 

Finally, to clearly resolve these Lagrangian transport structures, we consider the purely advective (kinematic) transport of fluid tracer particles in the absence of diffusion and dispersion. Although the presence of such dispersion alters solute transport from the purely advective case, to properly understand these transport structures it is necessary first to resolve them in the absence of dispersion, as is provided by the model statement above. Later studies will focus on how these structures control transport of dispersive species such as solutes, micro-organisms and colloids. Thus we maintain that our approach, although idealised, still is sufficiently representative of key dynamical interactions to provide useful conceptual information on transient groundwater discharge phenomena, and also to present a sound basis for future, more sophisticated investigations.

\section{Measures of Lagrangian complexity}
\label{sec:3}
Before embarking on the analysis of our spectral simulations we first introduce a set of convenient topological tools to gauge the Lagrangian complexity in flows. These tools all involve the calculation of flow paths for fluid elements within the flow field; these flow paths are solutions of the purely advective equation
\begin{equation} \label{eq:advection}
    \frac{d\mathbf{x}}{dt} = \mathbf{v}(\mathbf{x}(t),t) 
\end{equation}
\noindent with initial position of the fluid element $\mathbf{x}(t_{0}) \equiv \mathbf{X} = (x_{0},y_{0})$. By convention we usually set $t_{0} = 0$ for ease of comparing flow paths; any exceptions will be noted explicitly, e.g. see Figure \ref{fig:originplot} and related text.

\subsection{Poincar\'e sections}
\label{sec:poincare}
The Lagrangian histories of fluid elements in the periodic flow may be examined using a Poincar\'e section, which is a stroboscopic map of transport that is formed by releasing a large number of massless advecting particles in the flow field. At regular time intervals $\mathcal{P}$, the particle locations are recorded. Successive pictures are then assembled in one image (the Poincar\'e section) by superposing to provide a view of the evolution of the particle distribution as it advects through the domain under the influence of the subject flow. By choosing the Poincar\'e  period to be equal to the forcing period $P$, i.e. $\mathcal{P} = P$, the section discards intra-period particle displacements and focuses solely on the inter-period motions. In Appendix D in \cite{Trefry:2019aa}, we showed that whilst the fluid velocity field $\mathbf{v}(\mathbf{x},t)$ is, in general, not volume preserving (in that $\nabla\cdot\mathbf{v}\neq 0$) due to spatio-temporal fluctuations in the porosity field $\varphi$, the velocity field does satisfy the constraint
\begin{equation}
    \det \mathbf{F}(t,t_0;\mathbf{X})=\exp\left[\int_{t_0}^t \nabla\cdot\mathbf{v}(\mathbf{x}(t,t_0;\mathbf{X}),t^\prime)dt^\prime\right]=\frac{\varphi(\mathbf{X},t_0)}{\varphi(\mathbf{x}(t,t_0;\mathbf{X}),t)},\label{eq:deformation}
\end{equation}
for a fluid tracer particle with position $\mathbf{x}(t,t_0;\mathbf{X})$ at time $t$ that was initially at Lagrangian position $\mathbf{X}$ at time $t_0$. In equation (\ref{eq:deformation}), $\mathbf{F}\equiv d\mathbf{x}/d\mathbf{X}$ denotes the fluid deformation gradient tensor, which evolves as
\begin{equation}
    \frac{\partial\mathbf{F}}{\partial t}=\nabla\mathbf{v}(\mathbf{x}(t,t_0;\mathbf{X},t)^\top\cdot\mathbf{F}(t,t_0;\mathbf{X}),\,\,F(t=t_0,t_0;\mathbf{X})=\mathbf{1},\label{eqn:Fevolve}
\end{equation}
where $\det\mathbf{F}=1$ corresponds to volume-preserving deformation of fluid elements. As the porosity field $\varphi$ is time-periodic with respect to the characteristic period $P$, then for any integer $k$, (\ref{eq:deformation}) simplifies to
\begin{equation}
    \det \mathbf{F}(t_0+k P,t_0;\mathbf{X})=\frac{\varphi(\mathbf{X},t_0)}{\varphi(\mathbf{x}(t_0+n P,t_0;\mathbf{X}),t_0+k P)}=\frac{\varphi(\mathbf{X},t_0)}{\varphi(\mathbf{x},t_0)},\label{eq:pdeformation}
\end{equation}
hence fluid deformation around $k$-periodic points (denoted $\mathbf{X}_{k,p}$, where $\mathbf{x}(t_0+k P,t_0;\mathbf{X}_{k,p})=\mathbf{X}_{k,p}$) in the stroboscopic Poincar\'{e} map for $\mathcal{P}=P$ is volume-preserving:
\begin{equation}
\det \mathbf{F}(t_0+k P,t_0;\mathbf{X}_{k,p})=1.    
\end{equation}
As shown in Figure~\ref{fig:psharmonicstriped}, this structure renders the $\mathcal{P}=P$ Poincar\'{e} map everywhere \emph{continuous} in that superposed material lines do not overlap in this stroboscopic map, whereas Poincar\'{e} periods $\mathcal{P}$ that are not integer multiples of the characteristic period $P$ result in overlapping superposed material lines, rendering identification of coherent Lagrangian structures difficult. As discussed in Section~\ref{sec:2}, one feature of multimodal forcing is that the characteristic period $P$ can be significantly longer than that of the individual modes, hence the residence times of fluid particles may only extend to several multiples of $P$. This means that coherent ($\mathcal{P}=P$) Poincar\'{e} sections under multimodal forcing may be sparser than those for single-mode forcing (Figure~\ref{fig:psharmonicstriped}), hence additional visualization tools may be required to fully resolve the Lagrangian transport structure of these flows.

We consider two distinct types of Poincar\'e section that highlight different aspects of the transport structure, termed \emph{distributed} and \emph{inlet} Poincar\'e sections. The distributed Poincar\'e section is formed by advecting a set of tracer particles whose initial locations are distributed throughout the aquifer domain, whereas the inlet Poincar\'{e} section is formed from an initial distribution of points along the inland (right) boundary and allowed to advect under the regional flow gradient $J$ toward the tidal discharge boundary. Such Poincar\'e sections are useful for summarising topological controls (which reflect the presence of periodic points, manifolds etc) on the regional discharge dynamics in a graphical manner. However it is also useful to develop quantitative metrics by which the Lagrangian structures embodied in such topological descriptions may be compared.

\subsection{Elliptic point metric ($\mathcal{E}$)}
Periodic points $\mathbf{X}_{k,p}$ are critical in organizing the Lagrangian kinematics of time-periodic flows~\citep{Ottino:1989aa}, but locating all such points is not trivial in heterogeneous and time-dependent flows. These periodic points may be classified in terms of their local net fluid deformation over the periodicity of the point (quantified by $\mathbf{F}(t_0+k P,t_0;\mathbf{X}_{k,p})$), which in 2D may arise as either a circulation (elliptic) or a saddle-type (hyperbolic) deformation, respectively according to whether the traces $\textrm{tr}\mathbf{F} \equiv \textrm{tr}(\mathbf{F})$ satisfy $|\text{tr}\mathbf{F}(t_0+k P,t_0;\mathbf{X}_{k,p})|>2$ or $|\text{tr}\mathbf{F}(t_0+k P,t_0;\mathbf{X}_{k,p})|<2$~\citep{Ottino:1989aa,Ottino:2004aa}. Elliptic points are associated with closed circulation regions (known as Kolmogorov-Arnol'd-Moser (KAM) tori) that trap material elements indefinitely, whereas hyperbolic points generate exponential stretching of material elements, and can lead to chaotic dynamics~\cite{Aref:1984aa,Ottino:1989aa}.  

Quasi-periodic systems \citep{Wiggins:1992aa,Kapitaniak:1994aa}, as may be generated by anharmonic $M$-spectra (see Appendix \ref{app:period}), require more sophisticated techniques to identify and map periodic points. For example, in our context, the period of an anharmonic spectrum varies as $1/\delta\omega$ where $\delta\omega$ is the smallest difference between the modal frequencies in the forcing spectrum. As $\delta\omega \to 0$ the characteristic period $P \to \infty$, hence periodic points may be difficult to fix in the presence of a regional flow strong enough to ensure that the mean regional flow discharges in a time less than $P$. However, based on the analysis of \citet{Kapitaniak:1994aa}, we anticipate that the impact of periodic points located in a 1-spectrum problem may persist to some degree in an associated anharmonic 2-spectrum problem as long as the perturbing amplitude of the second spectral component is low. In general, formal identification and analysis of quasi-periodic manifolds (as discussed by \citet{Wiggins:1992aa}) is beyond the scope of this paper.

A grid search algorithm has been developed to locate elliptic points in a periodic velocity field by repeated particle tracking. This approach relies on the finding of \citet{Wu:2019aa} that flow complexity is indicated efficiently by the presence or absence of elliptic points, i.e. that elliptic points are a convenient surrogate measure of Lagrangian complexity, specifically trapping phenomena. Elliptic points are important from a transport perspective as these points are associated with ``trapped regions'' of the flow that involve closed particle trajectories (KAM tori) that cannot exist in heterogeneous Darcy flow in the absence of sources and sinks or boundary effects~\citep{Bear:1972aa}. Initial testing shows that this brute force algorithm can identify discrete elliptic points (but not cantori or stochastic regions) separated in space by a distance of at least one correlation length $\lambda$. The grid search step is expressed as $\epsilon \, \lambda$ where $0 < \epsilon < 1$ measures the resolution of the grid search; typically $\epsilon = 0.25$ is employed here. We can use this grid search algorithm to define a non-negative elliptic point metric, $\mathcal{E}$, which counts the number of distinct elliptic points in a flow field $\mathbf{v} = \mathbf{v}(\mathbf{x},t)$. An $\mathcal{E}(\mathbf{v})$ value of zero corresponds to unbifurcated flow with only regular, open particle trajectories, whereas $\mathcal{E}(\mathbf{v}) > 0$ indicates the presence of topological bifurcation(s) to closed or chaotic trajectories characteristic of Lagrangian complexity. Larger values of $\mathcal{E}$ indicate more complex Lagrangian flows with multiple bifurcations.

\subsection{Finite-time Lyapunov exponent (FTLE)} \label{sub:ftle}
While elliptic point metrics readily assist us to classify flows as unbifurcated (open) or bifurcated (closed), and the presence of elliptic points also indicates the presence of an similar number of hyperbolic points (due to preservation of the Euler characteristic), they are unable to detect and quantify Lagrangian chaos which can be important for understanding fluid mixing in these systems. The finite-time Lyapunov exponent can quantify stretching and folding along individual flow paths, using the deformation gradient tensor approach as discussed in Appendix D of \citet{Trefry:2019aa}. Almost all flow paths in our system that originate at the inland boundary eventually discharge along the tidal boundary after some finite travel time, the exceptions being those that commence on a KAM orbit around an elliptic point (and hence have zero FTLE). Therefore, because of the finite paths, non-zero FTLE evaluations do not rigorously prove the existence of persistent chaos in our time-dependent open flows. Instead such evaluations will be taken to signify local (in space and time) stretching manifolds which nonetheless may control enhanced mixing zones in time-dependent discharge flows.

\subsection{Topological entropy and folding maps} \label{sub:folding}
Despite the utility of the FTLE to quantify enhanced mixing phenomena in a close neighbourhood of a flow path (via the deformation gradient tensor $\mathbf{F}$), assessing the effective Lyapunov exponent over many flow paths traversing a finite region in a flow field can be computationally expensive. As an alternative, estimation of a lower bound to the \textit{topological entropy} of a flow field has proven useful in analysing braiding/entanglement phenomena \citep{Thiffeault:2010aa} to constrain effective topological exponents over ensembles of flow paths.
The E-tec topological entropy method \citep{Roberts:2019aa,Tan:2019aa} is computationally efficient and readily generates estimates for the local entropic exponent from time-ordered, discrete positional samples $\textbf{x}_{i}$ from collections of neighbouring flow paths. E-tec fits a deformable bounding surface (an elastic ``band") to the initial locations of flow paths in the ensemble and then tracks the total count of edges of this elastic surface $W(t)$ (the band weight) as the ensemble of flow paths deforms and entangles; the topological entropy is approximated by computing the exponential growth rate $\Lambda_{\textrm{TE}}$ of $W(t)$ \citep{Roberts:2019aa}, i.e. $W(t) \sim \exp(\Lambda_{\textrm{TE}} \, t)$. The exponent $\Lambda_{\textrm{TE}}$ provides an accurate lower bound to the rate of topological entropy generated by the trajectory ensemble. We will use the E-tec method to estimate ensemble topological entropic exponents $\Lambda_{\textrm{TE}}$ for entangling discharge flows. 

A different view of the topological structure of the flows is given by constructing \textit{folding maps}, which simply plot the $y$ values of the starting versus exiting locations of each flow path. For ensembles of flow paths this map provides clear information on the presence and magnitude of braiding and folding structures in the flow which, in turn, is diagnostic of entanglement and positive topological entropy.

\subsection{Residence time distributions (RTDs)}
Residence time distributions are important measures of fluid migration and fate in groundwater systems. The amount of time $(\tau)$ a fluid element spends within a groundwater flow field is a key factor in determining the potential for fluid alteration due to chemical and/or biological reaction as the fluid element moves along its flow trajectory in space and time, and hence RTDs are commonly used tools in contaminant hydrology. RTDs may also be used in a diagnostic manner to provide information on the Lagrangian topology of flows \citep{Mezic:1999aa}. For example \citep[see][]{Trefry:2019aa}, depending on the choice of starting positions within a flow, singular residence time values or intervals may indicate the presence of elliptic points and KAM orbits or even excluded zones, while discontinuous or rapidly varying RTDs may indicate entanglement of flow paths and/or hyperbolic points. Here we examine RTDs for two cases: (i) inlet RTDs providing information on how regional flows discharge to the tidal boundary, and (ii) distributed RTDs potentially identifying zones within the flow where trapping, segregation or holdup occurs. As shown in \cite{Trefry:2019aa} and discussed in subsection \ref{sec:poincare}, the structure of the latter spatially resolved RTD conforms with the Lagrangian structures identified in the Poincar\'{e} section. This correspondence is especially useful in the case of multimodal forcing as the spatially resolved RTD can be used to provide a more complete visualisation of the coherent Lagrangian structures due to the possible sparsity of the multimodal Poincar\'{e} sections. In recognition of the increased computational demands that accompany multimodal analysis, we employ an $N$-step Poincar\'{e} mapping scheme for the RTD calculations as described in Appendix \ref{app:Nstep_mapping}.

\begin{table}[ht!]
\caption{List of spectral simulation cases and elliptic point metrics, showing the unscaled mode weights $\textsl{g}_1$ and $\textsl{g}_2$ and the effective tidal compressibility $\mathcal{C}_{\textrm{eff}}$. While $\mathcal{T}_{\textrm{eff}}$ values alter between 1- and 2-spectral cases, the $\mathcal{G}_{\textrm{eff}}$ values are held constant across all cases and identical log $K$ fields are used throughout. Cases A1 and C1 correspond to the example problem discussed in \citet{Trefry:2019aa}. The simulated topologies are summarized by ({\scriptsize KAM},$\mathcal{E}$,{\scriptsize SC}) triplets, where {\scriptsize KAM} is the number of elliptic orbits found in the grid search, $\mathcal{E}$ is the number of distinct elliptic points identified through clustering analysis, and {\scriptsize SC} is the number of stochastic/cantori zones inferred.}
\linespread{0.4}
 \begin{tabular}{|c c c c c|} 
 \hline
 Case  & $m = 1$ & $m = 2$ & $\mathcal{C}_{\textrm{eff}}$ & ({\small KAM},$\mathcal{E}$,{\small SC})\\ [0.5ex] 
 \hline
harmonic & $\theta_1 = 0,\omega_1 = 2 \pi$ & $\theta_2 = \pi/3,\omega_2=4 \pi$ & & \\
 A1 & $\textsl{g}_1 = 1$ & $\textsl{g}_2 = 0$ & 0.5 & (19,5,1)\\
 A2 & $\textsl{g}_1 = 1$ & $\textsl{g}_2 = 0.05$ & 0.5 & (30,4,1)\\
 A3 & $\textsl{g}_1 = 1$ & $\textsl{g}_2 = 0.1$ & 0.5 & (38,4,1)\\
 A4 & $\textsl{g}_1 = 1$ & $\textsl{g}_2 = 0.5$ & 0.5 & (30,4,1)\\
 A5 & $\textsl{g}_1 = 1$ & $\textsl{g}_2 = 1$ & 0.5 & (21,3,0)\\ [1.3ex]
 B1 & $\textsl{g}_1 = 1$ & $\textsl{g}_2 = 0$ & 0.05 & (7,2,0)\\
 B2 & $\textsl{g}_1 = 1$ & $\textsl{g}_2 = 0.05$ & 0.05 & (6,2,0)\\
 B3 & $\textsl{g}_1 = 1$ & $\textsl{g}_2 = 0.1$ & 0.05 & (1,1,0)\\
 B4 & $\textsl{g}_1 = 1$ & $\textsl{g}_2 = 0.5$ & 0.05 & (0,0,0)\\
 B5 & $\textsl{g}_1 = 1$ & $\textsl{g}_2 = 1$ & 0.05 & (0,0,0)\\ [0.6ex]
 \hline
anharmonic & $\theta_1 = 0,\omega_1 = 2 \pi$ & $\theta_2 =  \pi/3,\omega_2=2.04 \pi$ & & \\
 C1 (=A1) & $\textsl{g}_1 = 1$ & $\textsl{g}_2 = 0$ & 0.5 & (19,5,1)\\
 C2 & $\textsl{g}_1 = 1$ & $\textsl{g}_2 = 0.05$ & 0.5 & (21,4,1)\\
 C3 & $\textsl{g}_1 = 1$ & $\textsl{g}_2 = 0.1$ & 0.5 & (19,4,1)\\
 C4 & $\textsl{g}_1 = 1$ & $\textsl{g}_2 = 0.5$ & 0.5 & (0,0,0)\\
 C5 & $\textsl{g}_1 = 1$ & $\textsl{g}_2 = 1$ & 0.5 & (0,0,0)\\ [1.3ex]
 D1 (=B1) & $\textsl{g}_1 = 1$ & $\textsl{g}_2 = 0$ & 0.05 & (7,2,0)\\
 D2 & $\textsl{g}_1 = 1$ & $\textsl{g}_2 = 0.05$ & 0.05 & (2,1,0)\\
 D3 & $\textsl{g}_1 = 1$ & $\textsl{g}_2 = 0.1$ & 0.05 & (0,0,0)\\
 D4 & $\textsl{g}_1 = 1$ & $\textsl{g}_2 = 0.5$ & 0.05 & (0,0,0)\\
 D5 & $\textsl{g}_1 = 1$ & $\textsl{g}_2 = 1$ & 0.05 & (0,0,0)\\ [0.6ex]
 \hline
\end{tabular}
\label{table:simlist}
\end{table}

\begin{figure}[ht!]
    \includegraphics[width=11.8cm]{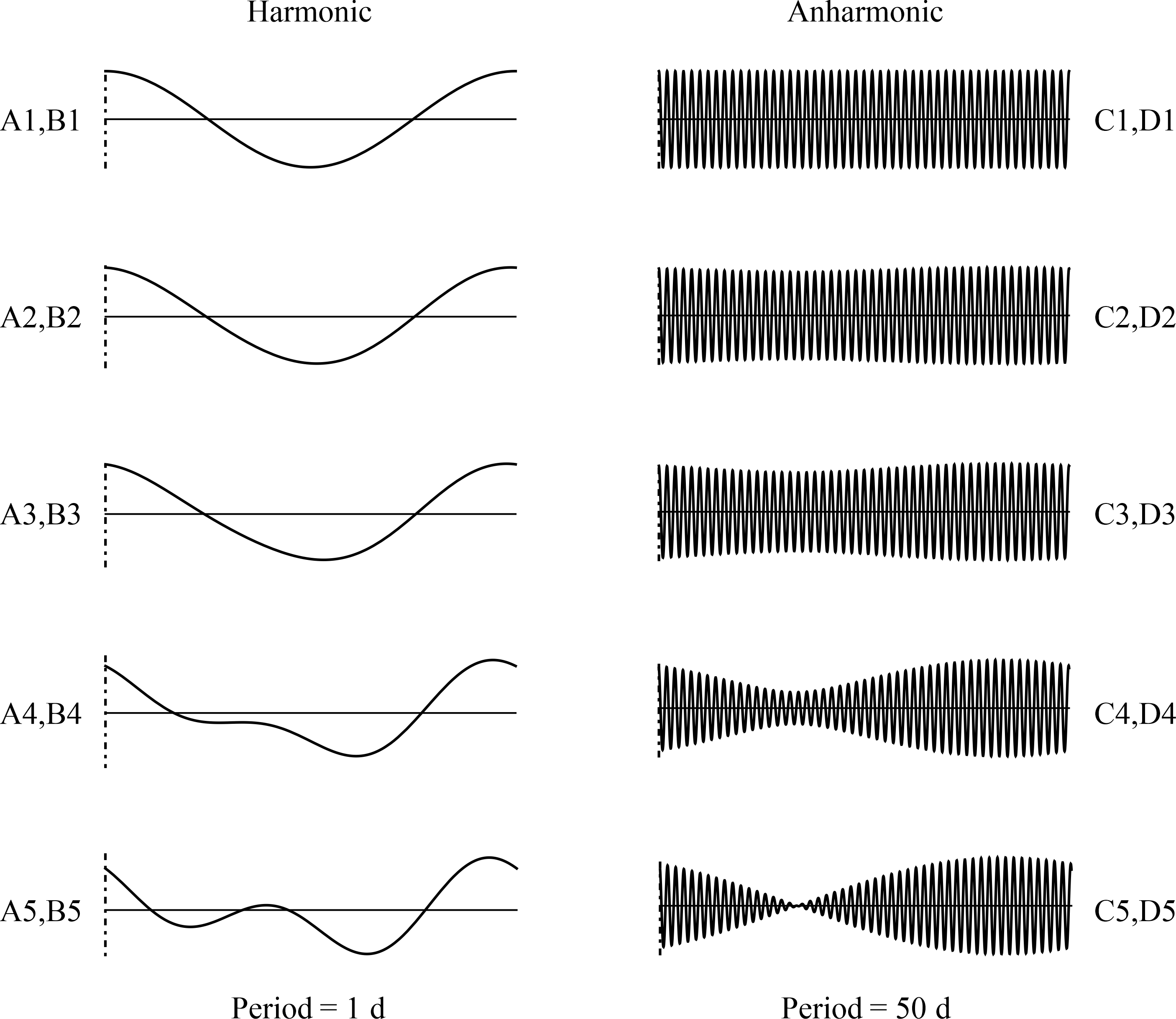}
    \caption{The tidal signals used as boundary conditions in the simulation cases of Table \ref{table:simlist}. Lettering denotes the simulation cases relevant to each signal and exact periods are shown below. Signals for cases A1,B1,C1,D1 are identical and purely sinusoidal (1-spectra), while the anharmonic signals display a strong quasi-periodicity of 1 day within an exact period of 50 days.}
    \label{fig:tidalsignals}
\end{figure}

\section{Spectral simulations and Lagrangian structure}
In this section we examine the effect of the several different kinds of forcing spectra upon the Lagrangian topology of the associated flow fields (see Appendix \ref{app:period} for more discussion on spectral terminology and conventions). Our approach is to run a series of simulations where the tidal signal spectrum is increasingly perturbed from a 1-spectrum to an equi-amplitude 2-spectrum for both harmonic and anharmonic cases. Where possible, physical parameters and log $K$ realisations are unchanged so that spectral effects are the sole focus. The only exception is a tenfold reduction in tidal compressibility $\mathcal{C}$ so that elastic compressibility effects on spectral phenomena can also be gauged across the simulations. Table \ref{table:simlist} lists the set of simulation cases considered here, and Figure \ref{fig:tidalsignals} depicts the different tidal boundary conditions used in the simulation cases, showing how the increasing amplitude of the higher frequency component perturbs the net tidal condition away from a pure sinusoid. By comparing simulation results we will be able to assess the influence of increasing spectral complexity on the Lagrangian structure of the calculated flow fields. 

\subsection{Spectral scaling approach}
At this stage we divert briefly to assess how best to ensure that comparisons between the different simulation cases in Table \ref{table:simlist} are meaningful given the variation of forcing spectra. Consider the case of a 2-spectrum with $\omega_{1} = \omega_{2}/2$, i.e. a diurnal fundamental mode $\omega_{1}$ and its (semidiurnal) first harmonic $\omega_{2}$ with varying amplitude $\textsl{g}_{2}$ (see left column of Figure \ref{fig:tidalsignals}). We commence with the trivial case A1 where $\textsl{g}_{2}=0$, and we choose problem parameters identical to the example of  \citet{Trefry:2019aa} so that the A1 case recovers that example precisely, including $\mathcal{C}_{\textrm{eff}} = 0.5$. In subsequent cases A2,A3,A4,A5 we simply vary the harmonic amplitude $\textsl{g}_{2}$, keeping $\mathcal{C}_{\textrm{eff}}$ and $\mathcal{G}_{\textrm{eff}}$ constant by scaling both the fundamental and harmonic amplitudes by $(\textsl{g}_{1}+\textsl{g}_{2})$, and evaluating effective quantities according to \eqref{eq:amplitudeweight}. The same scaling approach is used for the B,C,D cases; it is immaterial whether the second mode is harmonically related to the first mode or not. Appendix \ref{app:amplitudescaling} discusses an alternative scaling approach for $M$-spectra.

\begin{figure}[htp]
    \includegraphics[width=11.8cm]{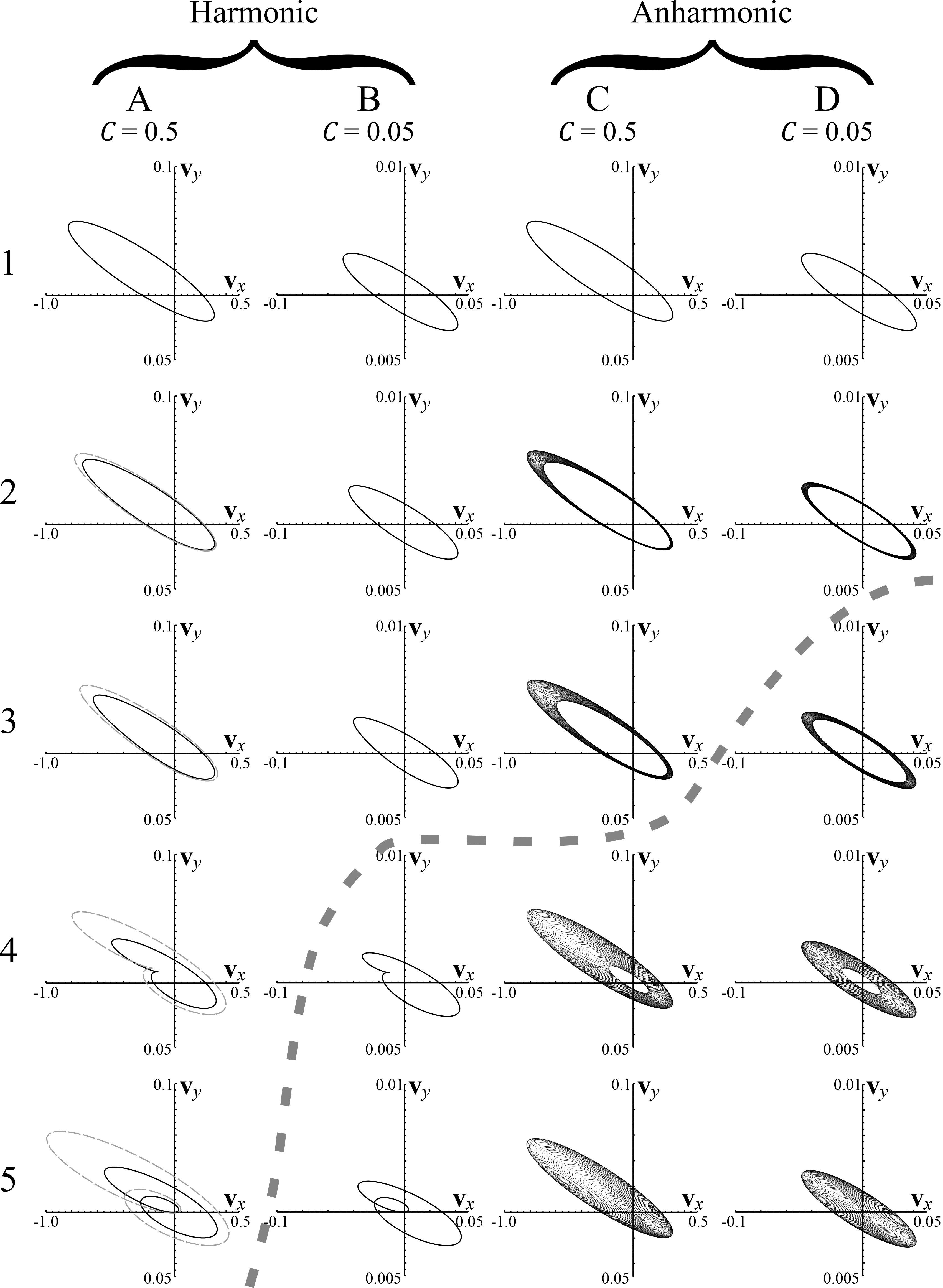}
    \caption{Velocity loci over one period $P$ for the simulation cases of Table \ref{table:simlist}, evaluated at an arbitrary location near the tidal boundary. The second modal amplitude ($\textsl{g}_{2}$) increases from top to bottom. The thick dashed grey curve divides simulation cases exhibiting $\mathcal{E} > 0$ (above) from those with $\mathcal{E} = 0$ (below). Cases A2-A5 also show loci calculated with power spectral weighting (thin dashed grey loci, see Appendix \ref{app:amplitudescaling} for details).}
    \label{fig:velocityloci}
\end{figure}

\subsection{Velocity loci}
We commence our survey of the simulation results by examining how the 2D local velocity loci traced by the two velocity vector components at a point over one characteristic period $P$ (see \citet{Trefry:2019aa} for a related discussion of flux ellipses) evolve as the harmonic amplitude increases in simulation cases A,B. In doing so we use an Eulerian view of the flow to understand what mechanics are operating at internal points in the domain. We calculate and compare velocity loci at a fixed internal point across all simulation cases, and the results are presented in Figure \ref{fig:velocityloci}. The figure shows that addition of the first harmonic perturbs the velocity locus of the fundamental mode, distorting the basic elliptic nature of the locus and eventually resulting in a complex, re-entrant curve as the amplitude of the harmonic reaches parity with the amplitude of the fundamental mode (cases A5 and B5). The figure displays velocity loci evaluated at an arbitrary interior point: similar complex velocity loci are not ubiquitous throughout the domain, but are prevalent throughout the zone of influence of the time-dependent forcing signal, i.e. for $x < x_{\textrm{taz}}$. 

The anharmonic cases (C,D) present a different morphology of velocity loci. In these cases, increasing $\textsl{g}_{2}$ amplitudes induce strong low-frequency beats with period $P = 50$. These beats modulate the effective amplitude of the tidal signal, causing the velocity loci to first thicken and eventually sweep out intervals as $\textsl{g}_{2}/\textsl{g}_{1} \to 1$. As shown by Figure \ref{fig:velocityloci} and Table \ref{table:simlist}, significant thickening of the anharmonic velocity loci for simulation cases C,D is sufficient to remove all elliptic points in the tidally active zone, for both high $\mathcal{C}$ and low $\mathcal{C}$ cases. As a whole, it is clear from the figure that formal elliptic points may variously be present or absent, depending on the nature of the forcing spectrum employed. 

\subsection{Trajectory sets}
\begin{figure}[htp]
    \includegraphics[width=11.5cm]{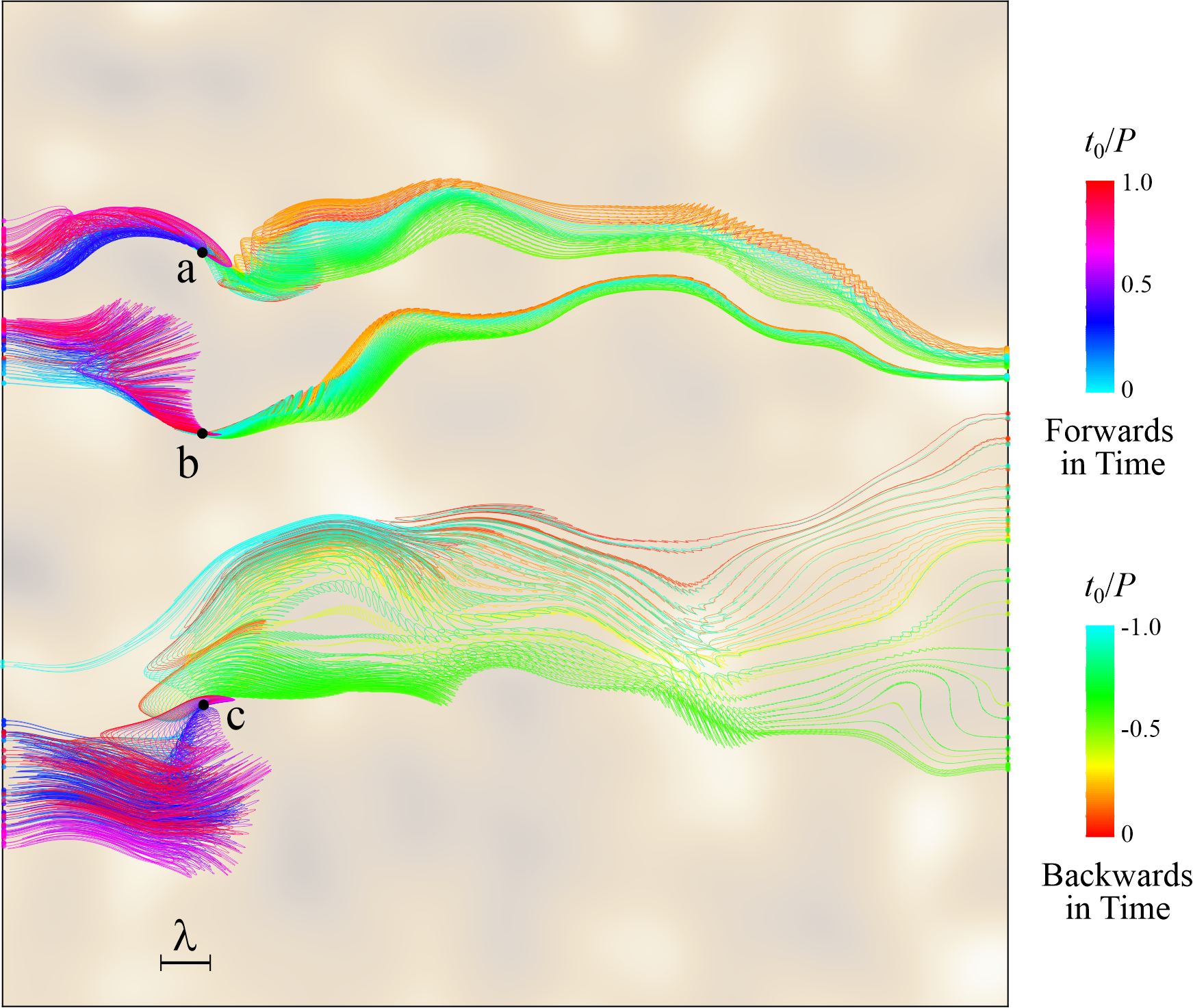}
    \caption{Forward- and backward-in-time temporal trajectory sets, evaluated from three arbitrary fixed locations a,b,c (black dots) for anharmonic simulation case C5. At each fixed location trajectory sets (for $-P<t_{0}<0$ and for $0<t_{0}<P$) are assembled from fifty flow paths commenced at different times ($t_{0}$) within the forcing period $P$. The brownish tones in the background describe the input log $K$ distribution. Note that location c samples fluid originating from both the inland boundary and from the tidal boundary (upper cyan flow paths) at different times within $P$.}
    \label{fig:originplot}
\end{figure}

Persisting with the Eulerian picture we consider forward- and backward-in-time trajectory sets originating at arbitrary fixed interior locations, as presented in Figure \ref{fig:originplot} for simulation case C5. Figure \ref{fig:originplot} demonstrates that, for any fixed location within the aquifer, local fluids sampled at different times within the forcing period $P$ may potentially have travelled to the sampling location from widely dispersed origins and, equally, may ultimately discharge to the tidal boundary at widely dispersed points. The structures of the trajectory sets are complex, with apparent folding and entanglement of flow paths and even bifurcation, as is shown in the backward-in-time trajectory set for location c which contains flow paths which originate from both the inland and tidal boundaries. The local stretching, folding and entanglement of fluid elements engendered by the complex velocity orbits presented in Figures \ref{fig:velocityloci} and \ref{fig:originplot} are indicative of enhanced mixing structures and may conveniently be studied and quantified by Lagrangian measures.

\subsection{Poincar\'e maps and material elements}
\begin{figure}[htp]
    \includegraphics[width=11cm]{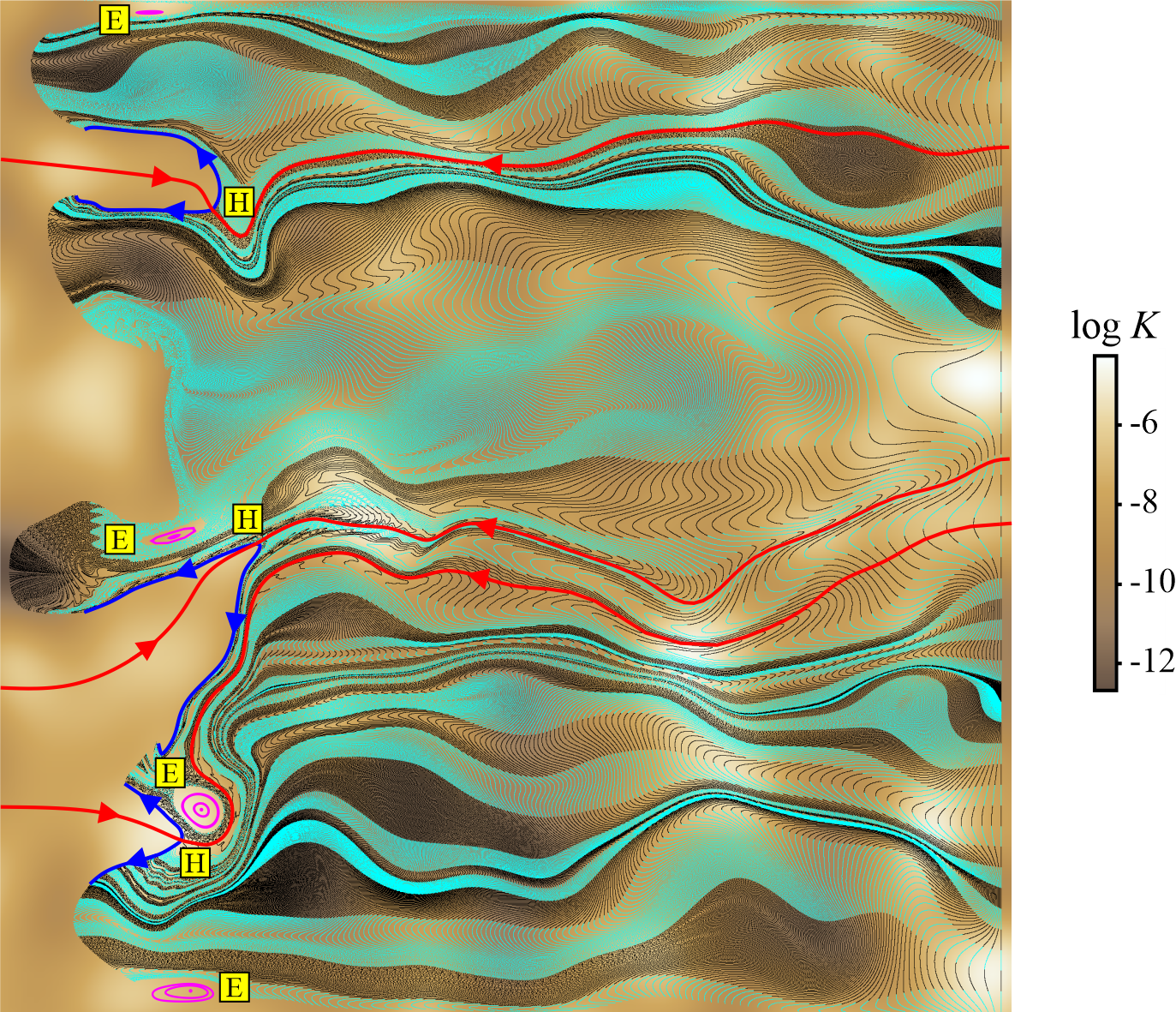}
    \caption{Poincar\'e section for the harmonic case A5. The Poincar\'e return period $\mathcal{P} = P$. Unstable (red) and stable (blue)  manifolds intersect at hyperbolic points (H). KAM tori (magenta) surround elliptic points (E). See text for explanation of cyan and black colouring. The uppermost KAM torus has small area and was not identified by the grid computation for $\mathcal{E}$ in Table \ref{table:simlist}.}
    \label{fig:psharmonicstriped}
\end{figure}

\begin{figure}[htp]
    \includegraphics[width=9.5cm]{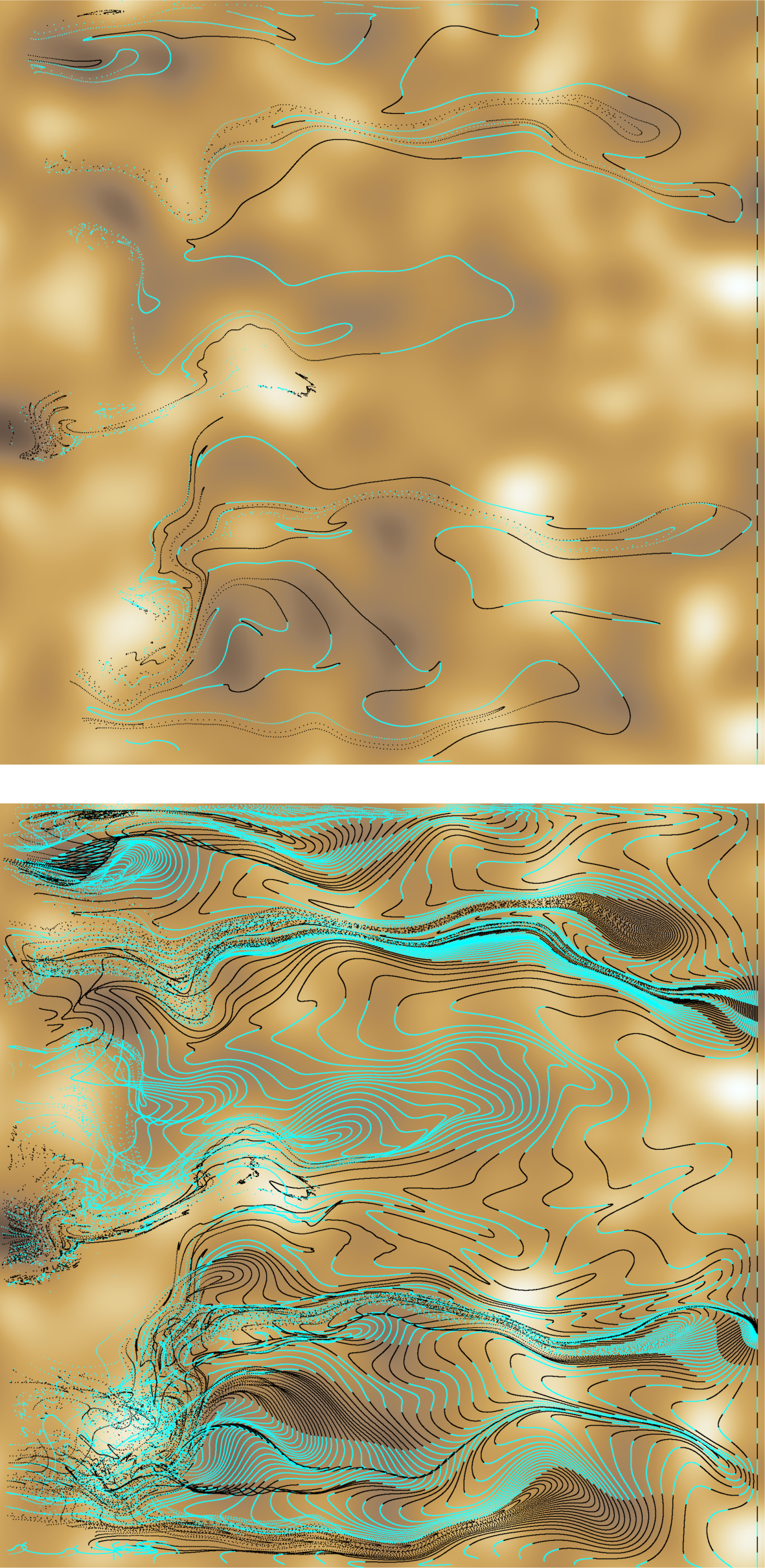}
    \caption{Poincar\'e sections for the anharmonic case C5. The Poincar\'e return period $\mathcal{P}$ takes the values $\mathcal{P} = P$ (top plate, volume-preserving) and $\mathcal{P} = P/10$ (bottom plate, non-volume-preserving). See text for explanation of cyan and black colouring.}
    \label{fig:psanharmonicstriped}
\end{figure}

The Lagrangian history of fluid elements in the flow may be studied using Poincar\'e sections, as shown in Figures~\ref{fig:psharmonicstriped} and  \ref{fig:psanharmonicstriped} for two values of the Poincar\'e return period $\mathcal{P}$. The sections are formed by releasing a large number (20,000) of tracer particles along the right-hand inland boundary, forming a dense line and allowing them to advect with the periodic flow. Additionally, we ascribe alternating colors (cyan and black) to groups of adjacent points along the initial line near the inland boundary, forming regular stripes which may be interpreted as linear material elements. The colour of each point is then maintained as it moves along its trajectory to the discharge boundary, with the result that the advective distortion of each coloured stripe (material element) becomes clear in the Poincar\'e section. Elements variously stretch, fold and deform as they move along their trajectories. 

In Figure \ref{fig:psharmonicstriped} for the harmonic simulation case A5, several coherent Lagrangian structures are clear, including two KAM islands (marked E) each encircling an elliptic point, several stable and unstable manifolds emanating from hyperbolic points (marked H), and a well-defined tidal emptying boundary \citep{Wu:2019aa} with pronounced inland-penetrating features. This figure is the harmonic counterpart of the Poincar\'{e} section in the example problem of \citet{Trefry:2019aa}, which itself is equivalent to the present simulation case A1. There is clear evidence of stretching and folding of material elements as they are increasingly influenced by the tidal boundary. The ensemble element stretch statistics are reported in Figure \ref{fig:rtdanharmonic}, showing a log-normal distribution and a maximum stretching factor of $O(10^{3})$.

In Figure \ref{fig:psanharmonicstriped} for the anharmonic simulation case C5, the top plate shows the continuous nature (non-overlapping material lines) of the Poincar\'e section formed using the exact forcing period $\mathcal{P} = P$, whereas the Poincar\'e section in the bottom plate is formed using $\mathcal{P} = P/10$ and is thus not a formally volume-preserving map, showing dense overlapping of material lines at different times within the forcing period. Because $P$ is large for this simulation case, the exact Poincar\'e section (top plate) is sparse and does not necessarily convey the full complexity of the flow. Equally, the bottom plate, while showing zones of complex overlapping of material elements, may also be misconstrued to suggest that material elements actually cross at the same instant in time: \textit{synchronous} crossing of material elements is not admitted by our Darcy flow, so the $\mathcal{P} = P/10$ Poincar\'e section should be examined with the understanding that it displays a superimposed ensemble of material elements with locations captured at different times within the forcing period. That said, element stretching by more than three orders of magnitude is present in this Poincar\'e section as well as numerous folds and curves, resulting in an approximately log-normal stretching histogram for the flow (see Figure \ref{fig:rtdanharmonic}). Regions with the strongest stretching/folding dynamics are spatially heterogeneous and are localized to apparent zones where intermittent inflow at the tidal boundary intermingles with the regional gradient flow, leading to the formation of recirculation zones and chaotic mixing.

\subsection{Elliptic points and trends}
We quantify the Lagrangian structures of the simulation cases in terms of the elliptic point metric $\mathcal{E}$. Calculated $\mathcal{E}$ values for the harmonic simulations (A,B) are presented in the rightmost column of Table \ref{table:simlist}. The results show that KAM orbits are found readily by the $\mathcal{E}$ grid search for all $\textsl{g}_{2}$ amplitudes in the harmonic A simulation cases. As $\textsl{g}_{2}$ increases the corresponding $\mathcal{E}$ estimate declines slowly, but the sole stochastic/cantori zone persists until case A5 ($\textsl{g}_{2} = 1$).

For the counterpart B simulation cases (see Table \ref{table:simlist}), corresponding to a reduced tidal compression ratio $\mathcal{C}_{\textrm{eff}} = 0.05$, numbers of KAM orbits and $\mathcal{E}$ values are smaller but mirror the A-simulations trend of declining as $\textsl{g}_{2}$ increases. The overall finding of reduced Lagrangian complexity for the low-$\mathcal{C}_{\textrm{eff}}$ simulations is in line with earlier studies \citep{Trefry:2019aa,Wu:2019aa} which showed that Lagrangian complexity is positively correlated with $\mathcal{C}_{\textrm{eff}}$, however the reducing trend of Lagrangian complexity $\mathcal{E}$ with increasing harmonic amplitude is a new result. This does not contradict the earlier findings because the present 2-spectrum results maintain constant $\mathcal{G}_{\textrm{eff}}$ as $\textsl{g}_{2}$ increases -- such scaling mechanisms are not catered for in the previous studies.

The results for the anharmonic simulations (C,D) display greater attenuation of elliptic point counts in all cases where $\textsl{g}_{2} > 0$, and elliptic points are absent for the larger $\textsl{g}_{2}$ amplitudes (see Table \ref{table:simlist}). The kinematic mechanisms controlling these elliptic point metrics are discussed in later sections, but we note that further insights may potentially be gained through the route of quasi-periodic analysis \citep{Wiggins:1992aa,Kapitaniak:1994aa}, especially for the anharmonic spectra. Further discussion of quasi-periodic approaches is beyond the scope of this paper.

\begin{figure}[htp]
    \includegraphics[width=9.5cm]{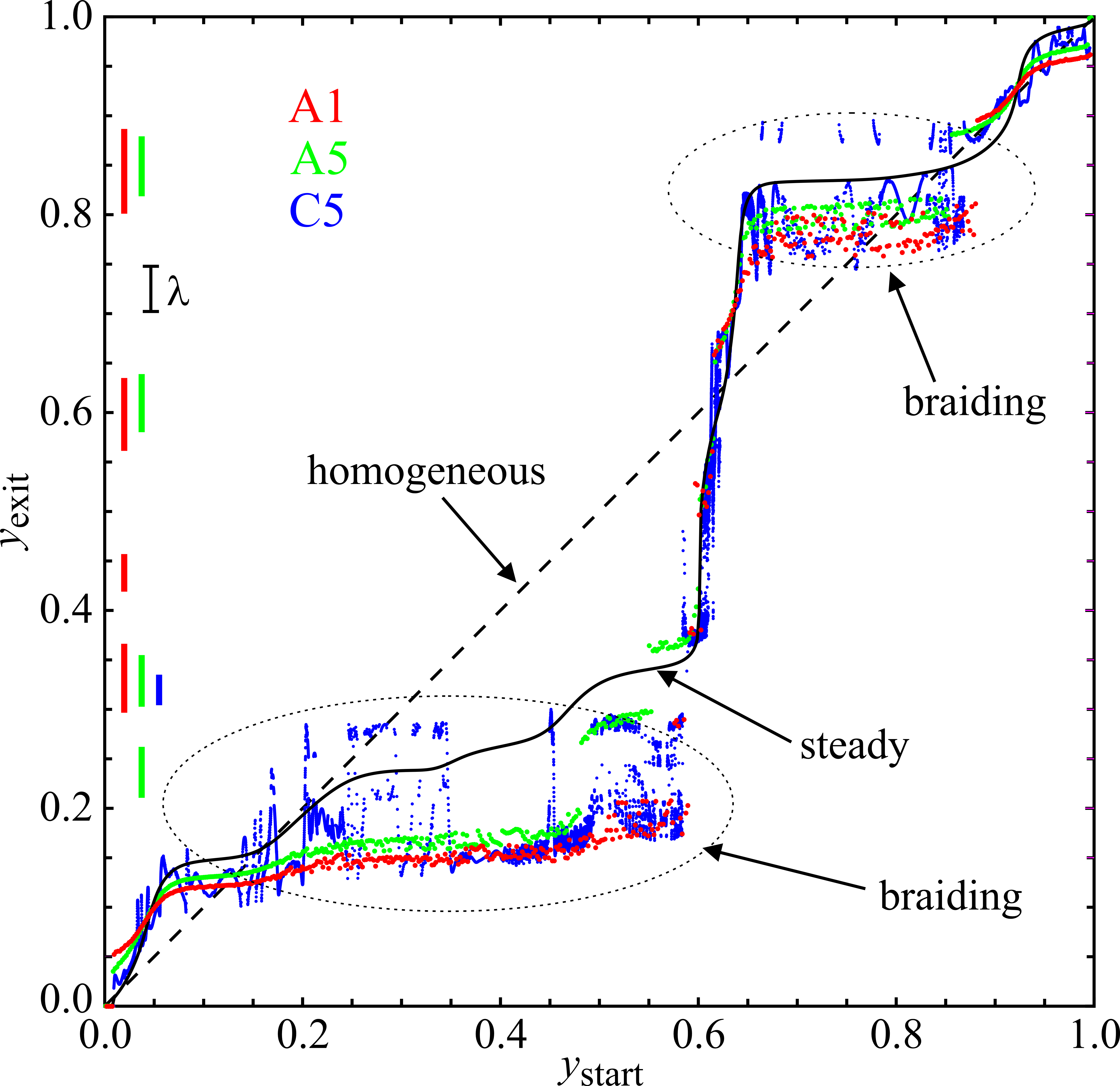}
    \caption{Folding maps for some of the simulation cases of Table \ref{table:simlist}. The homogeneous result is the 1:1 line (dashed); also shown is the folding map for the steady sub-problem (black curve) showing the effect of the log-$K$ heterogeneity. Maps are shown for spectral simulations A1 (red dots), A5 (green dots) and C5 (blue dots). Jagged or rough parts of the folding maps indicate braiding/entanglement of (initially) neighbouring flow paths. Excluded zones are indicated along the $y_{\textrm{exit}}$ axis by coloured bars.}
    \label{fig:foldingmap}
\end{figure}

\subsection{Folding maps}
In order to examine more closely the topological mechanisms that underpin these elliptic metrics, we consider the braiding properties of the flows by means of folding maps as discussed in section \ref{sub:folding}. The maps are constructed by calculating an ensemble of flow paths, e.g. up to 20,000 flow paths, with starting points distributed along the inland boundary and exiting at the tidal boundary. For each flow path the coordinate pair $(y_{\textrm{start}},y_{\textrm{exit}})$ is formed and the ensemble of pairs is plotted to form the associated folding map. Figure \ref{fig:foldingmap} shows three selected results from the simulation cases (A1,A5,C5), plus maps for the trivial cases of a homogeneous log-$K$ field, and for the steady sub-problem $\textsl{g}_{1} = \textsl{g}_{2} = 0$. This figure clearly shows that the folding maps are non-monotonic for the presented cases A1, A5, C5; this result obtains for all simulated cases. Non-monotonicity of the folding map indicates braiding/entanglement of flow paths. The folding map for anharmonic case C5 (blue data in Figure \ref{fig:foldingmap}) shows that the braiding amplitudes of neighbouring flow paths extend over large intervals of the discharge boundary equivalent to several $\lambda$ correlation scales. The braiding is dense for all simulation cases, but the braiding amplitude is least for harmonic and low-$\mathcal{C}$ simulations. Thus our results show that braiding/entanglement, which was reported in the 1-spectrum study of \citet{Trefry:2019aa}, is omnipresent in the 2-spectrum simulations and is enhanced by increasing amplitude of the second mode, i.e. increasing spectral complexity enhances topological complexity. 

The figure also shows that imposition of a forcing spectrum leads to clear but localised departures from the steady folding map (solid black curve). We interpret these departures, which correspond to the annotated braiding zones in Figure \ref{fig:foldingmap}, as examples of the dynamic accommodation between intruding tidal flows and discharging regional flows (see also Figure~\ref{fig:originplot}). At times during the forcing period the boundary heads are higher than those of the discharging regional flows and fluid intrudes into the aquifer domain. In some locations the local $K$ values are high enough to allow the intruding fluid to reside in the interior for several periods $P$, forming boundary recirculation zones that are visible in Poincar\'e sections, e.g. Figure \ref{fig:psharmonicstriped} and leading to exclusion zones (presenting as data gaps along the $y_{\textrm{exit}}$ axis in the folding maps, see coloured bars in  Figure \ref{fig:foldingmap}). Exclusion zones are rare for anharmonic forcing (C5), but harmonic forcing (compare A1,A5) does not diminish this recirculation/exclusion phenomenon. Nevertheless both harmonic and anharmonic forcing significantly enhance the braiding and entanglement of discharging flow paths, with anharmonic spectra making the stronger effect.

\begin{figure}[ht]
    \includegraphics[width=11.8cm]{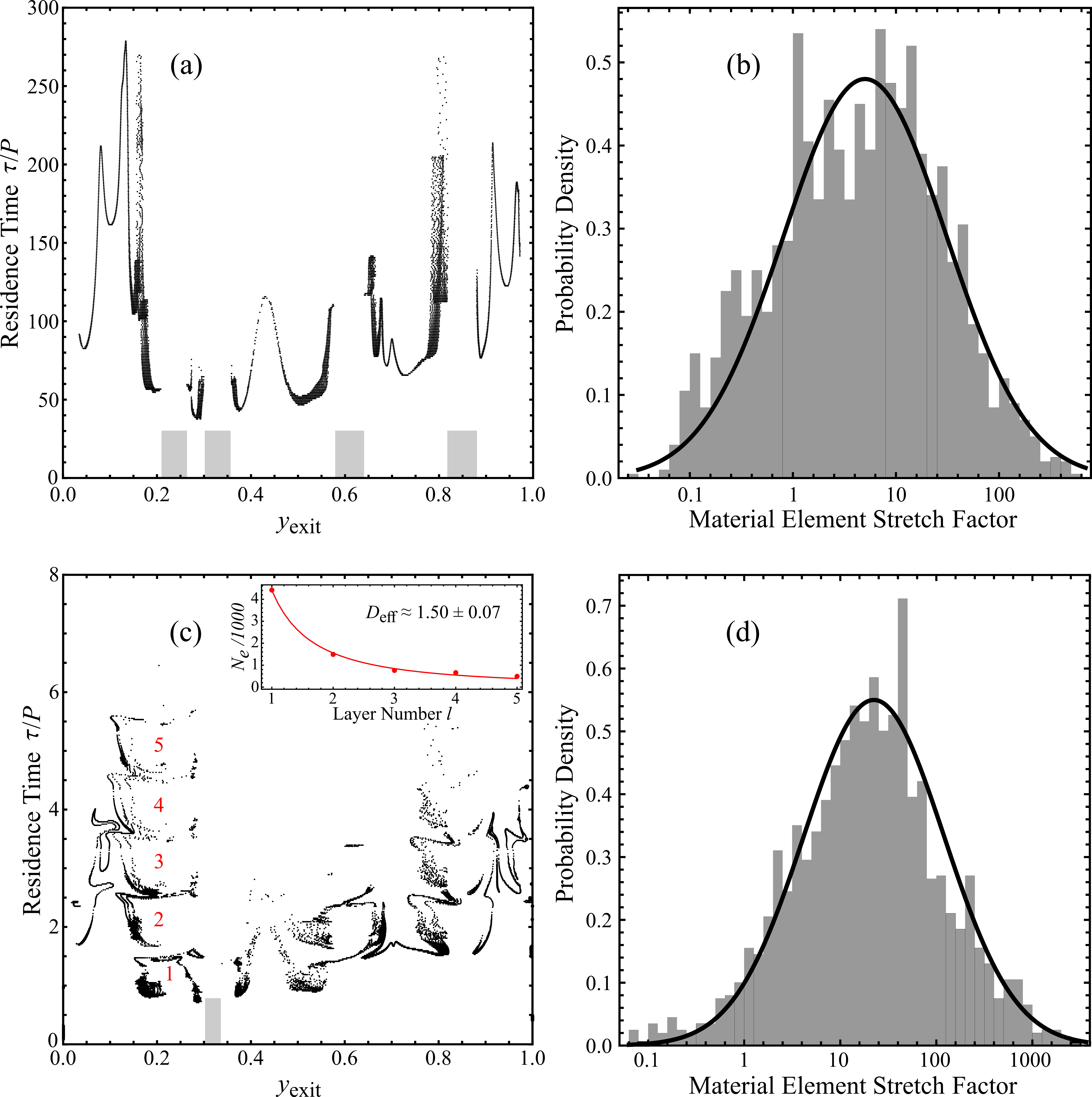}
    \caption{RTDs (left column) and material element stretch factor histograms (right column) for the harmonic case A5 (top row) and anharmonic case C5 (bottom row). The folding structures noted in Figure \ref{fig:foldingmap} are also evident in the RTDs. Exclusion zones are indicated by grey bars along the $y_{\textrm{exit}}$ axis in plots (a) and (c). The inset on the C5 RTD plot (c) shows the power law fit to the layered structures 1-5 lettered in red on the RTD. The histograms are approximately log-normal as shown by the superimposed Gaussian densities (b and d, black curves, fitted by eye).}
    \label{fig:rtdanharmonic}
\end{figure}

\subsection{Multivalued RTDs}
\label{rtdcantori}
The net effects of these complex transport phenomena are expressed most clearly in the RTDs for discharging regional flow. Figure \ref{fig:rtdanharmonic} plots the RTD for simulation cases A5 and C5 with a resolution of 20,000 flow paths. The figure shows the astounding result that high media compressibility leads to a multivalued RTD with folding/braiding structures \textit{discharging at the same location at the tidal boundary} at times separated by up to hundreds of tidal periods (A5) or five tidal periods (C5). This type of multivalued behaviour is also indicated in other simulation cases, e.g. A1, albeit with potentially lower residence time intervals, as evidenced in the folding maps of Figure \ref{fig:foldingmap}. Other features are noticeable in both A5 and C5 RTDs, including small but appreciable intervals of $y_{\textrm{exit}}$ where no RTD points exist. As mentioned in the previous section, these data gaps were identified with \textit{exclusion zones} \citep{Trefry:2019aa} where regional discharge is locally prevented by focused net inflow from the tidal boundary, often leading to landward-penetrating zones culminating in hyperbolic points due to the formation of recirculation flows.

The layered structures in the left column of Figure \ref{fig:rtdanharmonic} contain  fractal signatures associated with leakage of trajectories from a chaotic saddle in an open flow. Consider the layered structures numbered 1,2,3,4,5 above $y_{\textrm{exit}} = 0.2$ in part (c) of Figure \ref{fig:rtdanharmonic}, each delayed by one period with respective to the previous layer. Each one of these layers is associated with iterated material stretching and folding within a single chaotic saddle (hence they have the same basic structure). The longer persistence of trajectories that lie asymptotically close to the unstable manifold of this saddle accounts for the winnowing of trajectories with increasing layer number $l$. Counting the total number of points $N_{e}$ in each structure allows a decreasing curve to be plotted as a function of layer number $l$ (see inset). This curve obeys a power law $N_{e}(l) \sim l^{-D_{\textrm{eff}}}$ where $D_{\textrm{eff}}$ may be identified with the effective fractal dimension of a non-hyperbolic chaotic scattering process and is also related to the topological entropy generation rate of the saddle~\citep{Tel2005}. The estimated value is $D_{\textrm{eff}} = 1.50 \pm 0.07$ (standard error). Such non-hyperbolic scattering indicates the presence of a chaotic saddle in the flow characterised by transport barriers (cantori) which act to delay and mix fluid particles inside the saddle before eventual release (so-called \textit{stickiness}) according to a fractal density. A second layered structure with four layers exists in the same RTD at $y_{\textrm{exit}} = 0.8$, but the numbers of points are lower and the power law fit is less statistically significant. The smaller period $P$ for the harmonic case A5 yields an RTD that is very heavily layered and finely folded over small intervals of $y_{\textrm{exit}}$, making the chaotic scattering analysis infeasible at the present sampling resolution.

This result is important since it shows that dynamical traps exist in the C5 flow that are different to the intact elliptic islands detected by the elliptic point metric grid search reported in Table \ref{table:simlist}. This is because the dynamical features leading to stickiness may be exceedingly small, or even may have zero measure \citep{Tel2005}, yet their influences on the flow topology may be profound. Typically, as the level of Lagrangian complexity of a flow field increases, KAM islands and associated cantori around elliptic points shrink in size, but more elliptic and hyperbolic points arise in the Lagrangian topology, the latter of which generate chaotic saddles and stochastic layers. For this reason the elliptic metrics $\mathcal{E}$ reported in Table \ref{table:simlist} are best regarded as indicators of \textit{macroscopic} Lagrangian structures, and provide only coarse lower bounds to the true numbers of Lagrangian features in the flow that may induce holdup and mixing.

The histogram plots (parts b and d of Figure \ref{fig:rtdanharmonic}) show that the anharmonic signal generates one order of magnitude higher mean and peak stretching factors than does the harmonic signal, which may support the notion that anharmonic spectra promote fluid mixing more than harmonic spectra. However this conclusion is not supported by the Lagrangian observations presented in the following subsections, including Lyapunov exponents (Figure \ref{fig:lyapunovtraces}) and topological entropies (Figure \ref{fig:teresults}) where the presented data shows that the magnitudes of the exponents for the harmonic and anharmonic cases are similar. This apparent discrepancy may arise from the whole-of-flow nature of the stretching histograms as compared to the small ensembles used in the calculations of Lyapunov and entropic measures.

\begin{figure}[ht]
    \includegraphics[width=11.8cm]{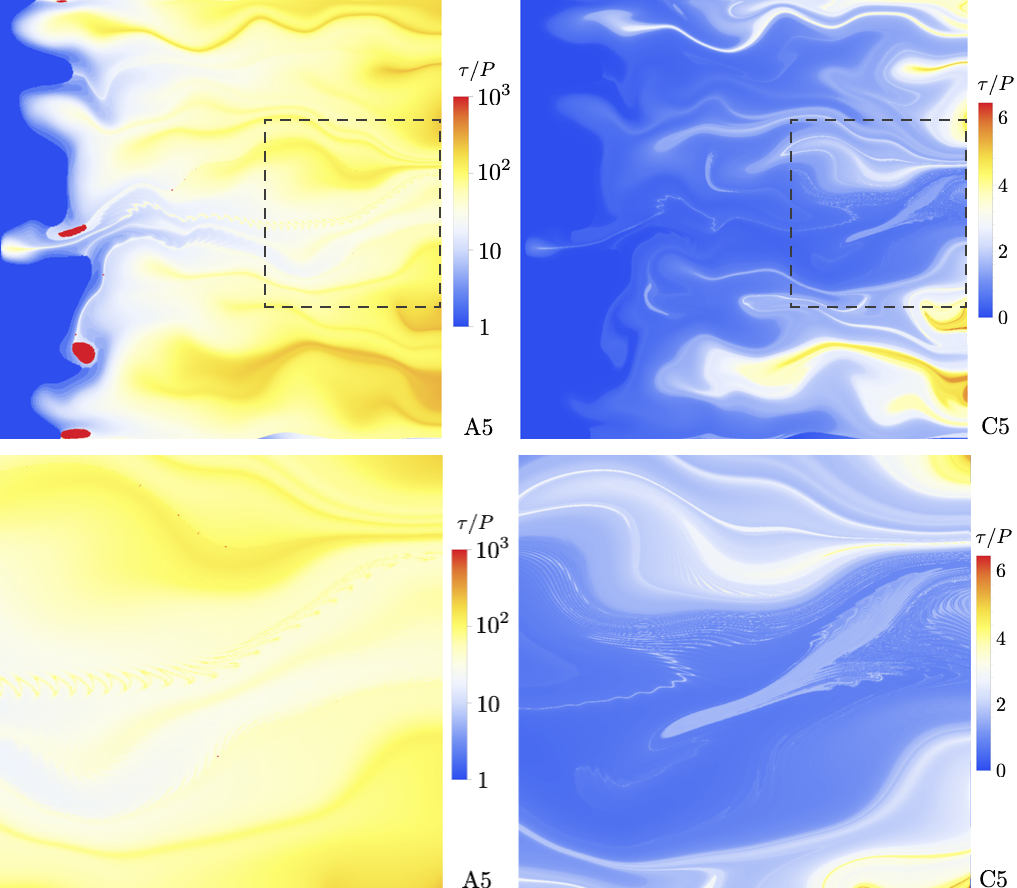}
    \caption{Spatially resolved RTDs for (left) A5 and (right) C5 cases over entire aquifer domain (top) and subdomain (bottom), as indicated by grey dashed line in top plots. Residence time is computed from initial position within the aquifer. Note RTD is given as a log scale for A5, but is linear for C5.}
    \label{fig:rtddistributed}
\end{figure}

Figure~\ref{fig:rtddistributed} illustrates spatially resolved distributed RTD plots for cases A5 and C5, where the residence time at any point is computed for a fluid particle trajectory with initial position at that location. The structure of the A5 and C5 RTD plots corresponds with the associated Poincar\'{e} sections shown respectively in Figures~\ref{fig:psharmonicstriped} and \ref{fig:psanharmonicstriped}, where the KAM tori manifest as trapped regions with unbounded residence times. Whilst there does not appear to be any evidence of the folded discharge dynamics shown in Figure~\ref{fig:rtdanharmonic}(c) in the top plots in Figure~\ref{fig:rtddistributed}, the more detailed bottom plots in Figure~\ref{fig:rtddistributed} show evidence of complex spatial RTDs for case C5. These complex distributions for case C5 are generated by chaotic saddles that are located near the tidal emptying boundary, but manifest upstream of these Lagrangian structures due to the scattering of particle trajectories as they pass through the aquifer domain. Conversely, folding of RTDs is significantly tighter for case A5 in Figure~\ref{fig:rtdanharmonic}(a); evidence of these dynamics is difficult to discern for case A5 in Figure~\ref{fig:rtddistributed}.

\begin{figure}[htp]
    \includegraphics[width=9.5cm]{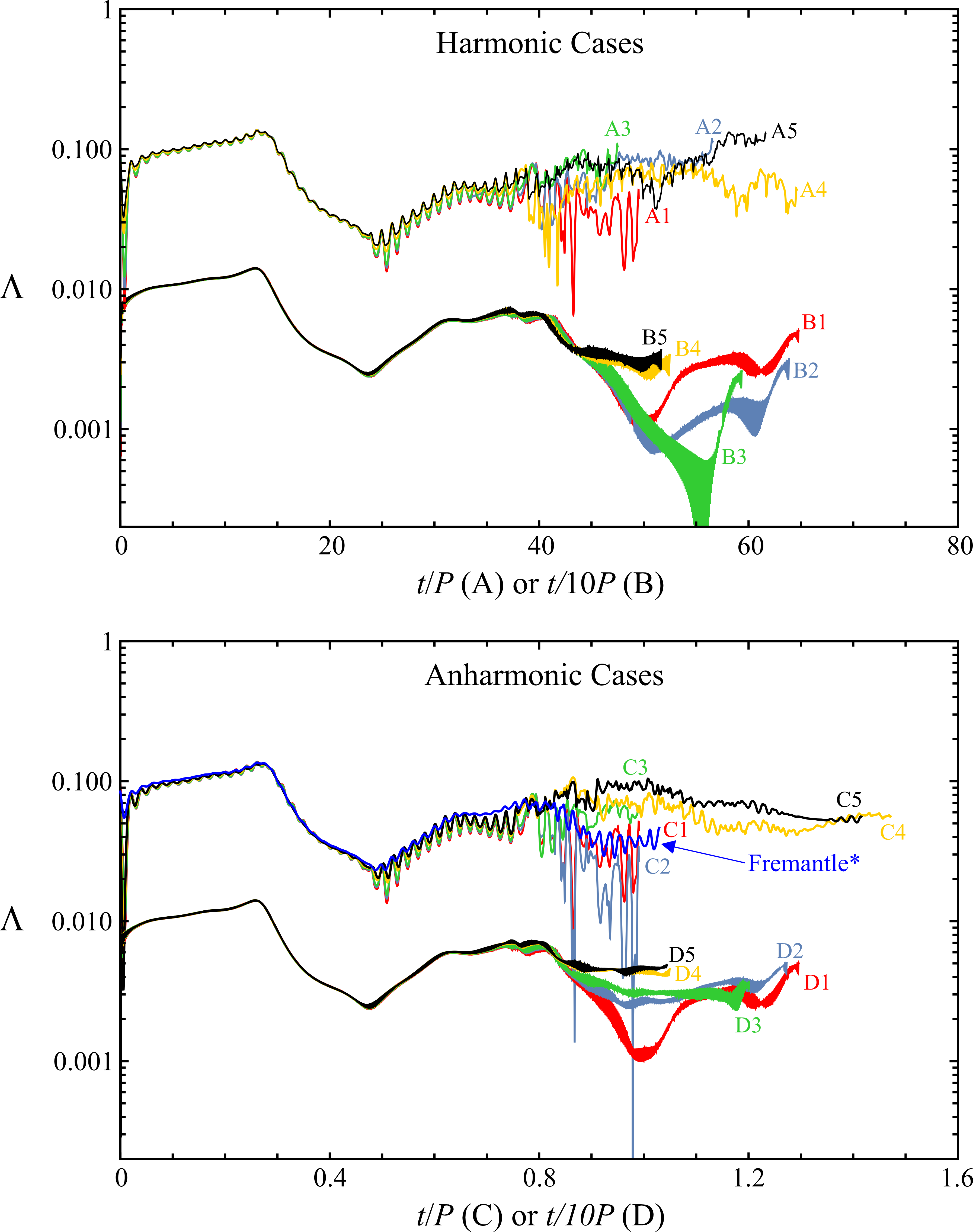}
    \caption{Finite-time Lyapunov exponents ($\Lambda$) plotted as functions of travel time for a single arbitrary flow path in the simulation cases of Table \ref{table:simlist}. The flow paths all have a common starting location near the inland boundary. The Fremantle spectrum has a period $P=100$ days and the associated $\Lambda$ trace is plotted against $2t/P$.}
    \label{fig:lyapunovtraces}
\end{figure}

\subsection{FTLE analysis}
Further information on the effect of increasing spectral complexity on chaotic signatures is presented in Figure \ref{fig:lyapunovtraces}, where a single starting location that generates flow paths near an unstable manifold is selected for FTLE analysis. The 1-spectrum results (cases A1, B1, C1, D1) are drawn in red, and show that $\Lambda$ values indicate significant stretching of fluid elements for A1 and C1 (high $\mathcal{C}$), and approximately tenfold weaker $\Lambda$ values for B1 and D1 (low $\mathcal{C}$). Addition of the second frequency mode leads to comparable or higher $\Lambda$ values in all 2-spectrum simulation cases. Therefore, based on these results, increasing spectral complexity does not necessarily reduce chaotic exponents and, more likely, often leads to increasing chaos (or increasing potential for chaos) in these open flows.

\begin{figure}[htp]
    \includegraphics[width=11.8cm]{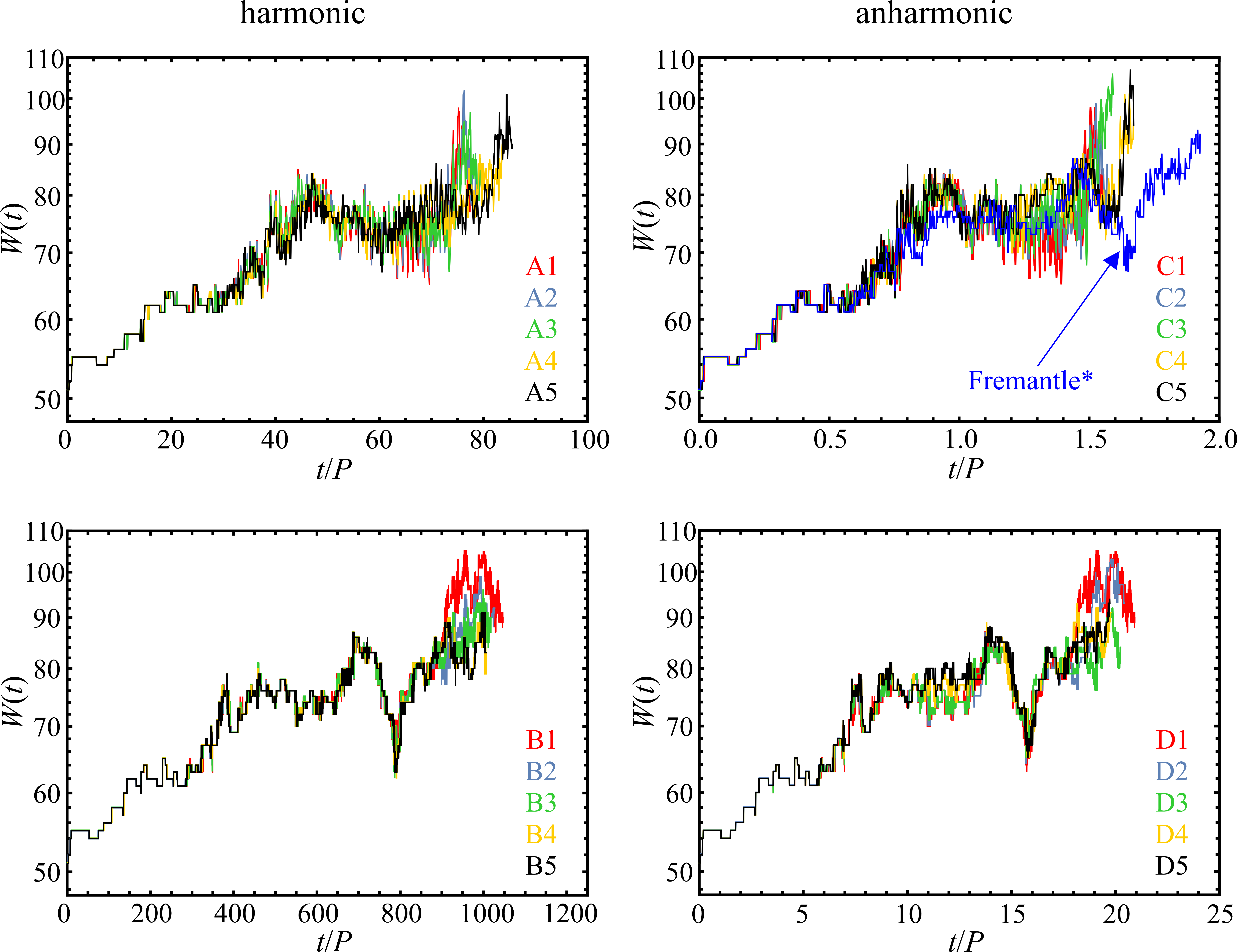}
    \caption{Total band weights $W(t)$ as functions of travel time $t$ generated via E-tec \citep{Roberts:2019aa} from identical ensembles of twenty paths for the simulation cases of Table \ref{table:simlist}. The top two plates are for the high tidal compressibility scenarios $(\mathcal{C}=0.5)$ and the lower two plates are for the low tidal compressibility scenarios $(\mathcal{C}=0.05)$. * The Fremantle spectrum has a period $P=100$ days and the associated band weight trace is plotted against $2t/P$.}
    \label{fig:teresults}
\end{figure}

\subsection{Topological entropy}
The entanglement of flow paths in our tidal flows generates increasing topological entropy with travel time, as shown in Figure \ref{fig:teresults}. Here the E-tec band weights are plotted as functions of time of flight along each trajectory for a small ensemble of twenty flow paths, identical for all simulation cases. As the flight time increases each fluid particle encounters variations in local conductivity and vorticity values. As the fluid particles near the tidal boundary, these variations increase in magnitude, leading to strong changes in the relative positions of the particles in the ensemble at each recorded time step. The overall effect across all simulation cases is to provide positive FTLE traces for all flow paths, and a net increasing trend for the ensemble band weight profiles $W(t)$, which is consistent with a positive topological entropy exponent $\Lambda_{\textrm{TE}}$. Mean $\Lambda_{\textrm{TE}}$ values for the simulation sets A, B, C, D are 0.0046, 0.00039, 0.25, 0.02, respectively, expressed in units of $P^{-1}$. Factors of ten between the results for A versus B, and C versus D are accounted for by the assumed reference porosity values $\varphi_{\textrm{ref}}$ used in the various cases.

\section{Example - Fremantle tide, Western Australia}
\label{sec:fremantle}
The results of the suite of idealised 2-spectrum simulations presented in the previous section have established that complex Lagrangian structure (stretching/folding of fluid elements, positive Lyapunov  and topological entropic exponents, braided residence time distributions) persists with the addition of a second tidal mode, harmonic or anharmonic, to the forcing spectrum. The only metric of Lagrangian complexity that declines significantly with increasing amplitude of the second mode is the number of elliptic points located by the grid search method and then only for anharmonic 2-spectra (cases C,D), due to the very long signal periods $P$ that arise where the anharmonicity satisfies $(\omega_{2}-\omega_{1})/\omega_{1} \ll 1$, and also due to the fact that some complex structures may be smaller than the grid search increment (as discussed in Section \ref{rtdcantori}). Nevertheless, as natural tidal forcings often contain many more than two periodic modes \citep{Munk:1966aa,Yousefi:2013aa}, it is useful to consider how a more complex and  realistic forcing spectrum may influence flow characteristics. We choose to explore this aspect by employing the Fremantle tidal signal, as previously studied in a tidal groundwater context by \citet{Trefry:2004aa}.

\begin{table}[ht!]
\caption{Dominant spectral modes for the Fremantle tidal signal (anharmonic 4-spectrum). Frequencies $f = \omega/2\pi$ are quoted to 2 decimal places, thereby ensuring an exact period of $P = 100 \; \textrm{d}$. }
\linespread{0.4}
 \begin{tabular}{|c c c c|} 
 \hline
Mode & Frequency \textit{f} $(\textrm{d}^{-1})$ & Amplitude $\textsl{g}$ (m) & Phase $\theta$ (radians) \\
 O1 & 0.93 & 0.1065 & -0.030 \\
 K1 & 1.00 & 0.1979 & 0.010 \\
 M2 & 1.93 & 0.0546 & -0.032 \\
 S2 & 2.00 & 0.0325 & 0.007 \\
 \hline
\end{tabular}
\label{table:fremantlesignal}
\end{table}

Fremantle is a port in the south-west of Western Australia, in a region of relatively low-amplitude tidal flows (see Figure 7 of \citet{Wu:2019aa}). The dominant lunar and solar constituents were extracted by \citet{Trefry:2004aa} from a set of ocean level observations made at the Fremantle gauging station every 15 min over a 57 day period in 1998 and are listed in Table \ref{table:fremantlesignal}. The table shows that the lunar modes (O1 and K1) together represent two-thirds of the amplitude of the overall tidal signal but the two solar modes (M2 and S2) while smaller are not negligible. We apply this anharmonic 4-spectrum as the boundary condition in our tidal problem, using the high-compressibility parameter values employed in the C simulation cases. We stress that in doing so we are not seeking to perform a site-specific groundwater flow simulation - to do so would require knowledge of the spatial statistics of hydraulic conductivity, porosity and storativity for the Fremantle coast, all of which are lacking at the resolution required here. Instead we are simply applying the Fremantle tidal signal to our example problem parameter set for the high compressibility $\mathcal{C} = 0.5$ case; we do this in order to gauge the effect of a more realistic tidal signal and to compare with the 2-spectrum results of the previous section. Even so, there are some important differences to note. First, the frequency resolution is such that the exact minimum period for the Fremantle signal is $P = 100 \; \textrm{d}$, twice as long as the period for the C simulation cases. Secondly the peak amplitude for the Fremantle signal is smaller, yielding the effective tidal strength $\mathcal{G}_{\textrm{eff}} = 3.9$ as opposed to $\mathcal{G}_{\textrm{eff}} = 10$ for the C simulations. We have chosen not to rescale the Fremantle signal to match the tidal strength of the earlier cases.

Figure \ref{fig:lyapunovtraces} shows the FTLE calculated using the Fremantle tidal signal. Even though the effective tidal strength is low, the Fremantle tidal signal produces an FTLE trace that is within the range of FTLEs determined for the C simulation cases, and is consistent with the C1 trace, albeit with a different time coordinate due to the longer period. Similarly the E-tec band weight profile generated by the flow path ensemble influenced by the Fremantle tidal signal is also broadly consistent with the case C counterparts, as shown in Figure \ref{fig:teresults}. The inferred $\Lambda_{\textrm{TE}}$ value is $0.39 \; P^{-1}$, similar to the average entropic exponent for the C simulation cases, despite the lesser effective tidal strength of the Fremantle tidal signal.

\begin{figure}[htp]
    \includegraphics[width=11.8cm]{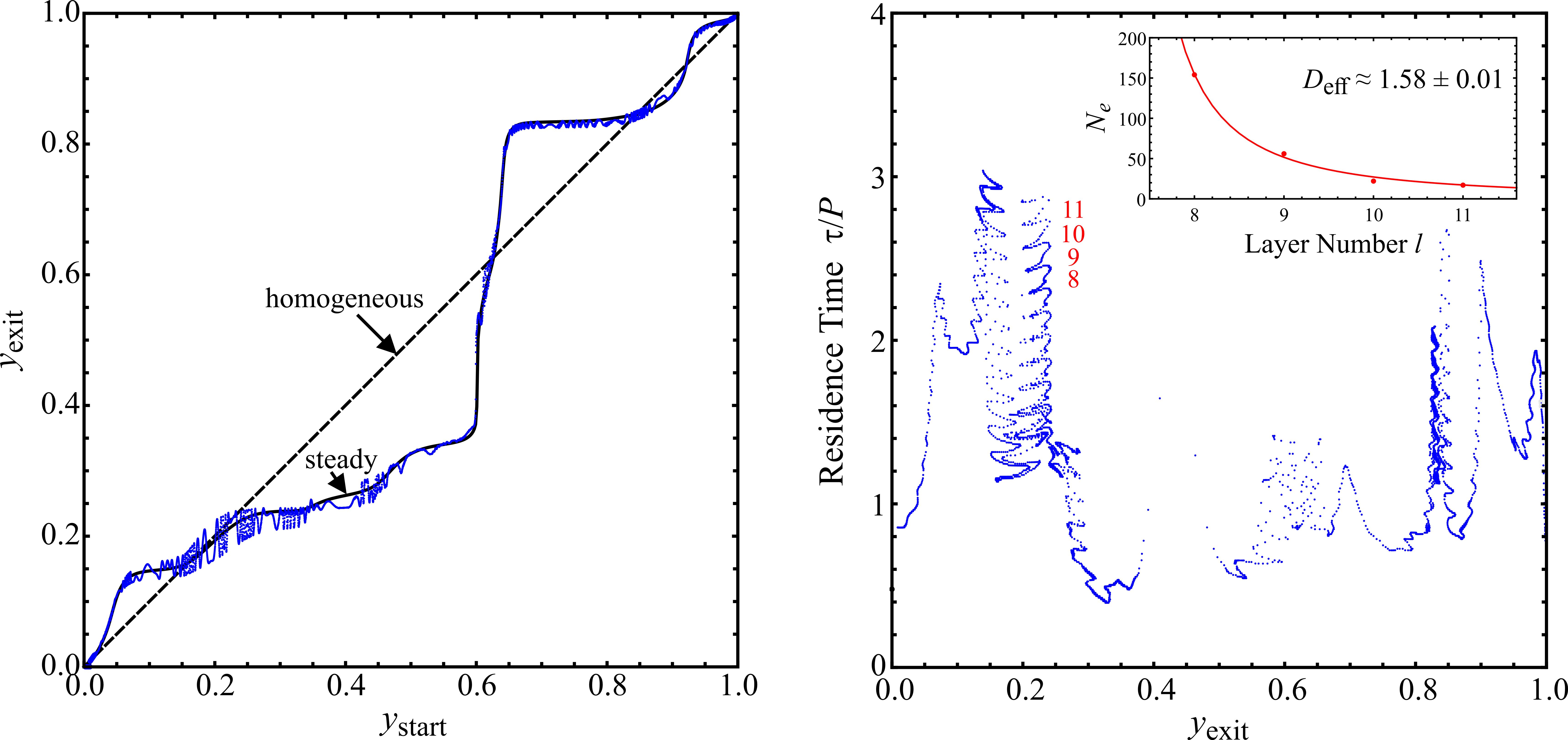}
    \caption{Folding map (left) and RTD (right) for the Fremantle tidal spectrum, calculated with 6000 points (blue). Despite the low $\mathcal{G}_{\textrm{eff}}$ value, the data shows significant folding structures in both plots. The inset on the RTD plot shows the power law fit to the layered structures 8-11 lettered in red on the RTD. (c.f. Figure \ref{fig:rtdanharmonic}). The sparsity of RTD points in the interval $y \in [0.4,0.5]$ indicates an exclusion zone \citep{Trefry:2019aa}.}
    \label{fig:fremantlefmrtd}
\end{figure}

 Figure \ref{fig:fremantlefmrtd} shows the folding map and RTD associated with the Fremantle tidal signal. Both of these display characteristics similar to the results of the previous simulation cases shown in Figures \ref{fig:foldingmap} and \ref{fig:rtdanharmonic}. For the folding map (Figure \ref{fig:fremantlefmrtd}, left) the data follow more closely the curve for the steady sub-problem than do the simulation cases, which is likely due to the low $\mathcal{G}_{\textrm{eff}}$ value. However, even with the lower tidal strength the folding map and the RTD (Figure \ref{fig:fremantlefmrtd}, right) still display a multivalued nature with multiple folding structures, suggesting that such structures are generic to multimodal spectra. In particular there are many layered folding structures in the RTD. These commonly overlap to various extents making particle counting in each layer difficult. Minimal overlap is observed for four layers annotated in red in Figure \ref{fig:fremantlefmrtd}; the resulting effective fractal dimension estimate of $D_{\textrm{eff}} = 1.58 \pm 0.01$ is similar to the value reported in Figure \ref{fig:rtdanharmonic}, indicating that non-hyperbolic scattering is also generated by the Fremantle spectrum. Taken together, all our results imply that increasing spectral complexity results in more complex Lagrangian structure. Certainly there is no evidence to suggest that a simple increase in spectral complexity intrinsically removes or reduces the potential for Lagrangian chaos in these time-dependent systems.

\section{Discussion}

With our survey of the simulation cases and the Fremantle tidal example complete, it is useful to list our findings and consider some ramifications for practical groundwater discharge investigations. Firstly, it is clear that natural seasonal and tidal forcings throughout the world are characterized by multimodal spectra ($M$-spectra) and that this must be taken into account when seeking to understand the subject groundwater dynamics. Prior studies \citep{Trefry:2019aa,Wu:2019aa} have shown that a rich set of dynamical phenomena can occur for groundwater flows discharging to boundaries governed by simple unimodal spectra. The present work has confirmed that these dynamical phenomena persist in the same groundwater systems when the boundary forcing is multimodal, and there is evidence to show that the Lagrangian complexity of the flow increases with the number of spectral modes.

Key features of the multimodal flows identified here include:
\begin{itemize}
  \item ubiquitous folding/braiding and bifurcations generated by quasi-periodic phenomena (Figures \ref{fig:originplot}, \ref{fig:foldingmap} and \ref{fig:fremantlefmrtd})
  \item chaotic and entropy-generating flows, quantified via FTLE and topological entropic measures (Figures \ref{fig:lyapunovtraces} and \ref{fig:teresults})
  \item multivalued residence time distributions consistent with transport barriers that hold up and mix fluid trajectories, releasing fractal assemblies of initial line segments (Figures \ref{fig:rtdanharmonic} and \ref{fig:fremantlefmrtd})
\end{itemize}

Even though the intensities of these complex Lagrangian structures decline with tidal compressibility $\mathcal{C}$, increasing spectral complexity can counterbalance this tendency and the potential ramifications for practical groundwater investigations are significant. In particular, the common practice of groundwater geochemical sampling to determine water quality indicators may need review for such systems. As is apparent from Figure \ref{fig:originplot}, and even if we ignore the possible presence of KAM islands and cantori, water samples taken at the same site but at different times within the spectral period $P$ may have originated from widely dispersed inland locations, or even from the discharge boundary itself. Equally, the ultimate discharge points of fluids released at the sample site at different times within $P$ may form a broad and possibly disjoint (bifurcated) distribution along the tidal boundary. In such a scenario the task of inferring representative and meaningful mean flow paths and transects to support geochemical transport analyses becomes difficult. As noted by \citet{Trefry:2019aa} for the unimodal case, the best defence against such confounding phenomena may be to ensure that the temporal resolution of sampling is high enough to detect any possible intra-period variations in water quality indicators at each distinct sampling location. If regular and reproducible variations are detected at any location then the need for a significantly more sophisticated sampling design to cater for complex Lagrangian dynamics may be indicated. Intra-period sampling techniques may thus also be required for other poroelastic contexts in geophysical industrial and biological systems.

Another feature of interest in the present results is the dynamic balance between regional discharge flow and the regular over-pressure of the tidal signal and how it leads to distinct, rapid discharge pathways in the regional flow system which make their way to the discharge boundary in channels of width $O(\lambda)$. These laminar outflow channels discharge at locations interspersed between more disordered zones where regional and boundary fluids strongly mix due to formation of recirculation zones, as identified by the strong folding dynamics observed in Figure~\ref{fig:foldingmap}. The location and geometry of the laminar-flow channels is difficult to predict just based on the macroscopic features of the log $K$ field alone, yet these few (dynamically controlled) channels are likely to govern the majority of regional solute mass transport as the outflow channels govern transport from the aquifer bulk to the tidal boundary, whereas the recirculation zones promote inflow of material from outside the tidal bound, and generation of strong local mixing between regional and boundary fluids. 

Finally we note that we have found evidence of elliptic points and/or cantori in nearly all simulations, either by direct calculation via the approximate elliptic point metric $\mathcal{E}$ (see Table \ref{table:simlist}) or indirectly via observation of fractal structures in residence time distributions (see Sections \ref{rtdcantori} and \ref{sec:fremantle}). It is well known that these features generate fluid segregation and delay phenomena in purely advective systems. More importantly for groundwater systems, where molecular diffusion is always present, these Lagrangian features can also modulate diffusive transport mechanisms for solutes leading to theoretical estimates of delays of hundreds or thousands of spectral periods \citep{Wu:2019aa}. Thus we should anticipate that complex Lagrangian structure, where present, will impact significantly upon solute transport and reaction in groundwater discharge systems. Quantifying these Lagrangian effects upon solute transport and reaction in groundwater systems (and other poroelastic systems) influenced by periodically forced boundaries is likely to be a fruitful and relevant area of research.

\section{Conclusions}

In this paper we have extended previous research into Lagrangian signatures of poroelastic Darcy flows influenced by sinusoidally forced boundaries to the case where the boundary forcing spectrum has multiple modes. Applying a range of Lagrangian tools to our open flow system, where an average regional flow gradient discharges to the forcing boundary, we have demonstrated that increasing spectral complexity results in a strengthening of Lagrangian structure, across a representative range of spectral types.

The mechanism for increased Lagrangian structure is the same as for single-mode flows; namely the establishment of complex (but still periodic) velocity loci in the tidally affected zone. These Eulerian features result in localised and transient velocity and vorticity bursts that distort Lagrangian flow paths, and providing repeated stretching and folding of fluid elements that is characteristic of chaotic advection. Complex spectral forcing also contributes to time-dependent entanglement of neighbouring flow paths, resulting in bifurcated discharges along the boundary and in strongly multivalued residence time distributions. Lyapunov and topological entropic exponents are generally positive, albeit, as expected, only weakly positive for simulation cases with low tidal compressibility values (all these phenomena vanish in the limit of zero compressibility). Fractal signatures have been detected in anharmonic simulations, indicating the presence of sticky manifolds arising from stochastic cantori and filamentary manifolds,  although direct observation of such features has not been achieved. Observations from an example study of the Fremantle tide in Western Australia, which is modelled as a 4-spectrum signal, indicates that increasing the number of tidal modes plays a greater role in generating complex transport dynamics than the specific amplitude of the individual modes.

The outcome of this study is a theoretical validation that multimodal spectral forcing acts to increase the Lagrangian complexity observed in prior (unimodal) works. This strengthens the case for complex Lagrangian structure to be present in coastal groundwater flows around the world. Discharging aquifers with the combination of high tidal compressibility $\mathcal{C}$, high log$K$ variance and multimodal boundary spectra appear to be strong candidates for observation of the rich set of dynamical phenomena discussed here, and, by extension, any periodically forced continua governed by similar poroelastic laws may also display complex Lagrangian structures.

\appendix
\section{Characteristic periods of commensurable $M$-spectra}
\label{app:period}
We wish to determine the characteristic period $P$ of a periodic boundary forcing condition $F(t)$ with discrete Fourier spectrum $\mathcal{F}$ given by

\begin{equation}
  \mathcal{F}(\omega) = \textsl{g}_{0} \; \mathbf{\delta}(\omega) + \sum_{m=1}^{M} \textsl{g}_{m}  e^{i\theta_{m}}  \mathbf{\delta}(\omega - \omega_{m})
\end{equation}

\noindent with amplitudes $\textsl{g}_{0} \geq 0$, $\textsl{g}_{m} > 0$, the phases satisfying $0 \leq \theta_{m} \leq 2\pi$, the commensurable modes $\omega_{m} > 0$ and $M$ finite. $\mathbf{\delta}$ is the Dirac delta function. Several cases are important.

\underline{Phase-split spectra:} Consider a 2-spectrum, i.e. a spectrum with two identical frequency modes $\omega_{1} = \omega_{2}$, with unequal phases $\theta_{1} \ne \theta_{2}$. Via phasor algebra it is easy to show that the sum of these two modes is equivalent to a single mode with well-defined amplitude $\big(  \textsl{g}_{1}^{2} + \textsl{g}_{2}^{2} + 2 \textsl{g}_{1} \textsl{g}_{2}\, \textrm{cos}[\theta_{1}-\theta_{2}]\big)^{1/2}$ and phase $\textrm{arctan}[(\textsl{g}_{1}\, \textrm{sin}\, \theta_{1} + \textsl{g}_{2}\, \textrm{sin}\, \theta_{2})/(\textsl{g}_{1}\, \textrm{cos}\, \theta_{1} + \textsl{g}_{2}\, \textrm{cos}\, \theta_{2})]$. Thus, by extension, any \textit{M}-spectrum of modes with identical frequencies can be expressed as a single mode with deterministic amplitude and phase. Hence, phase-split spectra are equivalent to the single-mode forcing scenario modelled earlier \citep{Trefry:2019aa,Wu:2019aa} and, as a consequence, in heterogeneous flows subject to boundary conditions represented by phase-split spectra no new topological structures can arise than those previously discussed.

Another consequence of the phasor analysis is that any \textit{N}-spectrum (phase-split or otherwise) can be reduced (deterministically) to an ordered sum of $M$ \textit{distinct} modes $(\textsl{g}_{m},\theta_{m},\omega_{m})$ for $m = 1,2,...,M$ where the strict inequalities $\omega_{1} < \omega_{2} < ... < \omega_{M}$ hold and $M \le N$. Thus we can restrict attention to \textit{M}-spectra where the frequency modes are distinct without loss of generality.

\underline{Indeterminate $P$:} Consider the trivial case where modes are of equal frequency and amplitude, and exactly out of phase, for example a 2-spectrum with $\omega_{1} = \omega_{2}$, $\textsl{g}_{1} = \textsl{g}_{2}$ and $|\theta_{1} - \theta_{2} | = \pi$. Then the sinusoidal terms cancel exactly and the spectrum $\mathcal{F} = \textsl{g}_{0} \; \mathbf{\delta}(0)$. Hence $F(t)$ is a constant and \textit{any real number} is formally a period of $F$. This trivial case is easily detected and avoided in practice.

\underline{Harmonic spectra:} Consider the commensurable case where the modes are ordered so that $\omega_{1} < \omega_{2} < ... < \omega_{M}$ and the \textit{fundamental mode} $\omega_{1}$ is an exact divisor of all modes. In this case all spectral modes are \textit{harmonically related} to $\omega_{1}$ and the characteristic (minimum) period of $F(t)$ is $P = 2\pi/\omega_{1}$.

\underline{Anharmonic spectra:} Consider the commensurable case where the modes are ordered so that $\omega_{1} < \omega_{2} < ... < \omega_{M}$ and $\omega_{1}$ is \textit{not} an exact divisor of all modes. In this case the spectral modes are \textit{anharmonically related} but a characteristic period can still be determined as follows (see \citet{Raykh:2017aa}). Form the list of modal periods $P_{modal} = \{2\pi/\omega_{1},2\pi/\omega_{2},...,2\pi/\omega_{M}\}$ and express each list member in rational form $p/q$ for integer values $p,q$. Multiply the $q_{m}$ to form the product $\Delta$ and determine the greatest common denominator GCD of the entries in the list $M = \Delta \times P_{modal}$. Calculate the least common multiple LCM of the list $M/\textrm{GCD}$. The characteristic period $P$ of $F(t)$ is then given by $(\textrm{GCD}/\Delta) \times \textrm{LCM}$. As an example, a 3-spectrum with modes $\omega_{1}=2\pi, \omega_{2}=2.1\pi, \omega_{3}=4\pi$ yields $P = 20$.

\section{Comments on spectral scaling}
\label{app:amplitudescaling}
We consider here the problem of scaling two different tidal spectra $\mathbb{S}_{M}$ and  $\mathbb{S}_{N}$ so that the effective problem parameters $\mathcal{C}_{\textrm{eff}},\mathcal{G}_{\textrm{eff}}$ (as defined in Table \ref{table:nondimpars}) are comparable between the two spectra. Noting that both $\mathcal{C}_{\textrm{eff}}$ and $\mathcal{G}_{\textrm{eff}}$ are both defined in terms of modal amplitudes $\textsl{g}_{m}$ it is natural to pursue a scaling of modal amplitudes. For example, if $M$-spectrum $\mathbb{S}_{M}$ contains modes  $(\alpha_{1},\theta_{1},\omega_{1}),(\alpha_{2},\theta_{2},\omega_{2}), ...,(\alpha_{M},\theta_{M},\omega_{M})$ then the sum of amplitudes for $\mathbb{S}_{M}$ is $a_{\mathbb{S}_{M}} = \alpha_{1} + \alpha_{2} + ... + \alpha_{M}$. Similarly, if $\mathbb{S}_{N}$ contains modes  $(\beta_{1},\phi_{1},\upsilon_{1}),\allowbreak(\beta_{2},\phi_{2},\upsilon_{2}),..., \allowbreak(\beta_{N},\phi_{N},\upsilon_{N})$ then the corresponding sum of amplitudes is $a_{\mathbb{S}_{N}} = \beta_{1} + \beta_{2} + ... + \beta_{N}$. In this picture, the rescaled spectrum $\mathbb{S}_{N}^{*}$ with modes $(\beta_{1} a_{\mathbb{S}_{M}}/a_{\mathbb{S}_{N}},\phi_{1},\upsilon_{1}),\allowbreak(\beta_{2} a_{\mathbb{S}_{M}}/a_{\mathbb{S}_{N}},\phi_{2},\upsilon_{2}), ..., \allowbreak(\beta_{N} a_{\mathbb{S}_{M}}/a_{\mathbb{S}_{N}},\phi_{N},\upsilon_{N})$ would yield peak amplitudes (and hence $\mathcal{C}_{\textrm{eff}},\mathcal{G}_{\textrm{eff}}$ values) equivalent to those generated by $\mathbb{S}_{M}$.

While the above amplitude scaling method is both appealing and simple, the kinematics generated from $\mathbb{S}_{M}$ and $\mathbb{S}_{N}^{*}$ contain a systematic bias, as illustrated in the first column of Figure \ref{fig:velocityloci}. In the Figure it is seen that as the amplitude of the first harmonic increases (down the A simulations column) the size of the velocity locus (solid curve) declines steadily, even though amplitude scaling is employed to keep $\mathcal{C}_{\textrm{eff}}$ and $\mathcal{G}_{\textrm{eff}}$ constant. The reason for this is that the waveform is increasingly being distorted by the first harmonic, reducing the number of peak amplitudes per period (see Figure \ref{fig:tidalsignals}). It is possible to use a spectral power scaling technique instead, where the metric of interest is the power spectral density $p_{\mathbb{S}_{M}}$ defined by
\begin{equation} \label{eq:powerspectraldensity}
    p_{\mathbb{S}_{M}} = \frac{1}{P} \int_{0}^{P} |\alpha_{1} e^{i (\omega_{1} t + \theta_{1})} + \alpha_{2} e^{i (\omega_{2} t + \theta_{2})} + ... + \alpha_{M} e^{i (\omega_{M} t + \theta_{M})}|^{2} dt
\end{equation}

\noindent where $P$ is the characteristic period of $\mathbb{S}_{M}$. For harmonic spectra, where all frequency modes are integer multiples of the fundamental $\omega_{1}$, the power spectral density reduces to $p_{\mathbb{S}_{M}} = (\alpha_{1}^{2} + \alpha_{2}^{2} + ... + \alpha_{M}^{2})/2$. The power-scaled spectrum $\mathbb{S}_{N}^{*}$ may be formed from modes $(\beta_{1} p_{\mathbb{S}_{M}}/p_{\mathbb{S}_{N}},\phi_{1},\upsilon_{1}),\allowbreak(\beta_{2} p_{\mathbb{S}_{M}}/p_{\mathbb{S}_{N}},\phi_{2},\upsilon_{2}),..., \allowbreak(\beta_{N} p_{\mathbb{S}_{M}}/p_{\mathbb{S}_{N}},\phi_{N},\upsilon_{N})$; this spectrum displays the same power spectral density as $\mathbb{S}_{M}$ but a different waveform. In particular, peak amplitudes will differ between the two spectra $\mathbb{S}_{M}$ and the power-scaled $\mathbb{S}_{N}^{*}$, leading to variations in $\mathcal{C}_{\textrm{eff}},\mathcal{G}_{\textrm{eff}}$ values. However, as seen from the dashed grey loci in the first column of Figure \ref{fig:velocityloci}, the velocity loci maintain approximately constant size as the amplitude of the first harmonic increases. 

It is not possible to deduce a scaling procedure that ensures that two different spectra generate identical tidal problem parameters $\mathcal{T}_{\textrm{eff}},\mathcal{C}_{\textrm{eff}},\mathcal{G}_{\textrm{eff}}$ and hence are directly comparable in terms of their Lagrangian structures. This difficulty makes comparison of simulation cases in Table \ref{table:simlist} and Figure \ref{fig:velocityloci} an imprecise process. Nevertheless, as we have discussed here, amplitude and power scaling both lead to topologically similar velocity loci at points internal to the problem domain, albeit with different magnitudes. The similarity of the velocity loci between the two scaling approaches provides confidence that our focus on comparing topological characteristics between simulation cases is reasonable. Apart from the limited results for power-scaled harmonic spectra presented in Figure \ref{fig:velocityloci}, amplitude scaling is employed by default in all simulations and results throughout this paper.

\section{$N$-step Poincar\'e mapping method}
\label{app:Nstep_mapping}

In \cite{Wu:2019aa} we described an efficient mapping method for advecting large numbers of fluid particles over one period $P$ of the flow. This method involved advecting a uniform square grid of tracer particles distributed over the entire aquifer domain for the duration of a single forcing period. The resultant particle locations were then used to construct the spline interpolation functions $(f_x,f_y)$ for the $x-
$ and $y-$coordinates at the end of a forcing period $(x_{n+1},y_{n+1})=(x((n+1) P),y((n+1) P))$ as a function of the coordinates $(x_n,y_n)=(x(n P),y(n P))$ at the start of the forcing period:
\begin{equation}
 x_{n+1}=f_x(x_{n},y_{n}), \quad y_{n+1}=f_{y}(x_{n},y_{n}).
\end{equation}
We use a similar method in this study to advect particles subject to multimodal forcing, with the exception that for cases with long characteristics periods we employ an $N$-step mapping method such that mapping is performed over a fraction $P/N$ (for some integer $N>1$) of the characteristic period $N$ to improve temporal resolution of RTDs, e.g. for anharmonic simulation case C5. In this case the $(x,y)$ coordinates at time $t=n P/N$ are denoted as $(x_{n},y_{n})=(x(n P/N),y(n P/N))$, and so the mapping to coordinates $(x_{n+1},y_{n+1})=(x(n P/N),y(n P/N))$ at time $t=(n+1) P/N$ is
\begin{equation}
 x_{n+1}=f_{x,n}(x_{n},y_{n}), \quad y_{n+1}=f_{y,n}(x_{n},y_{n}),    
\end{equation}
where the interpolated functions $(f_{x,n},f_{y,n})$ are constructed by integrating particles from the uniform square grid from time $t=n P/N$ to $t=(n+1) P/N$. Construction of $N$ such maps (where $n=1,2 ... N$) then allows rapid particle mapping over arbitrary time periods. The accuracy of the $N$-step map process was compared to the fully periodic case ($N=1$) and was found to be accurate to within the error of the fully periodic map.

\begin{acknowledgements}
The authors wish to thank Spencer Smith for his friendly assistance with the E-tec code. The Fremantle Harbour tidal data set used here was supplied by the Government of Western Australia, as originally acknowledged in \cite{Trefry:2004aa}.
\end{acknowledgements}

\bibliographystyle{spbasic}      
\bibliography{database.refTIPM}   

%
%

\end{document}